\documentclass{www2010-submission}
\usepackage{graphicx}
\usepackage{subfigure}
\usepackage{latexsym}
\usepackage{amsfonts}
\usepackage{amsmath}
\usepackage{amssymb}
\usepackage{url}
\usepackage{rotating}
\usepackage{color}
\usepackage{times}

\newcommand{\net}[1]{{\textsc{#1}}}

\newcommand{\denselist}{\itemsep -2pt\topsep-8pt\partopsep-8pt }

\begin{document}

\title{Empirical Comparison of Algorithms for\\Network Community Detection}

\numberofauthors{3}
\author{
\alignauthor
Jure Leskovec\\
\affaddr{Stanford University}\\
\email{jure@cs.stanford.edu}
% 2nd. author
\alignauthor
Kevin J. Lang\\
\affaddr{Yahoo! Research}\\
\email{langk@yahoo-inc.com}
% 3rd. author
\alignauthor
Michael W. Mahoney\\
\affaddr{Stanford University}\\
\email{mmahoney@cs.stanford.edu}
}

\maketitle

\begin{abstract}
Detecting clusters or communities in large real-world graphs such as large
social or information networks is a problem of considerable interest. In
practice, one typically chooses an objective function that captures the
intuition of a network cluster as set of nodes with better internal
connectivity than external connectivity, and then one applies approximation
algorithms or heuristics to extract sets of nodes that are related to the
objective function and that ``look like'' good communities for the
application of interest.

In this paper, we explore a range of network community detection methods in
order to compare them and to understand their relative performance and the 
systematic biases in the
clusters they identify. We evaluate several common objective functions that
are used to formalize the notion of a network community, and we examine several
different classes of approximation algorithms that aim to optimize such
objective functions. In addition, rather than simply fixing an objective and
asking for an approximation to the best cluster of any size, we consider a
size-resolved version of the optimization problem. Considering community
quality as a function of its size provides a much finer lens with which to
examine community detection algorithms, since objective functions and
approximation algorithms often have non-obvious size-dependent behavior.
\end{abstract}

\vspace{2mm} \noindent {\bf Categories and Subject Descriptors:}
H.2.8 {Database Management}: {Database applications -- Data mining}

\vspace{1mm} \noindent {\bf General Terms:} Measurement;
Experimentation.

\vspace{1mm} \noindent {\bf Keywords:} Community structure; Graph
partitioning; Conductance; Spectral methods; Flow-based methods.

\section{Introduction}
\label{sec:intro}
Detecting clusters or communities in real-world graphs such as large social
networks, web graphs, and biological networks is a problem of considerable
practical interest that has received a great deal of attention~\cite{newman02community,guimera04_fluctuations,Fortunato07_ResolutionPNAS,clauset04large_JRNL,KLN07_robustness_JRNL}.
A ``network community'' (also sometimes referred to as a module or cluster) is typically thought of
as a group of nodes with more and/or better interactions amongst its members
than between its members and the remainder of the
network~\cite{RCCLP04_PNAS,newman02community}.

To extract such sets of nodes one typically chooses an
objective function that captures the above intuition of a community as a set
of nodes with better internal connectivity than external connectivity.
Then, since the objective is typically NP-hard to optimize
exactly~\cite{Leighton:1988,Arora:2004,Schaeffer07_survey}, one employs
heuristics~\cite{newman02community,karypis98_metis,dhillon07graclus} or
approximation
algorithms~\cite{Leighton:1999,spielman96_spectral,andersen06local} to find
sets of nodes that approximately optimize the objective function and that
can be understood or interpreted as ``real'' communities.
Alternatively, one might define communities operationally to be the output of
a community detection procedure, hoping they bear some relationship to the
intuition as to what it means for a set of nodes to be a good
community~\cite{newman02community,newman04community}.
Once extracted, such clusters of nodes are often interpreted as organizational units in social
networks, functional units
in biochemical networks,
ecological niches in food web networks, or
scientific disciplines in citation and collaboration
%%%OLDCITE%%% networks~\cite{RCCLP04_PNAS,palla07_groupNature}.
networks~\cite{newman02community,RCCLP04_PNAS}.

In applications, it is important to note that heuristic approaches to and
approximation algorithms for community detection often find clusters that
are systematically ``biased,'' in the sense that they return sets of nodes with properties
that might be substantially different than the set of nodes that achieves
the global optimum of the chosen objective.
For example, many spectral-based methods tend to find compact clusters at
the expense that they are not so well separated from the rest of the
network; while other methods tend to find better-separated clusters that
may internally be ``less nice.''
Moreover, certain methods tend to perform particularly well or particularly
poorly on certain kinds of graphs, \emph{e.g.}, low-dimensional manifolds or
expanders.
Thus, drawing on this experience, it is of interest to compare these
algorithms on large real-world networks that have many complex structural
features such as sparsity, heavy-tailed degree distributions, small
diameters, etc. Moreover, depending on the particular application and the
properties of the network being analyzed, one might prefer to identify
specific types of clusters. Understanding structural properties of clusters
identified by various algorithmic methods and various objective functions can guide in selecting
the most appropriate graph clustering method in the context of a given network
and target application.

In this paper, we explore a range of different community detection methods in
order to elucidate these issues and to understand better the performance and
biases of various network community detection algorithms on different kinds
of networks. To do so, we consider a set of more than 40 networks; 12 common objective
functions that are used to formalize the concept of community quality; and 8
different classes of approximation algorithms to find network communities.
One should note that we are not primarily interested in finding
the ``best'' community detection method or the most ``realistic'' formalization
of a network community. Instead, we aim to understand the structural
properties of clusters identified by various methods, and then depending on the
particular application one could choose the most suitable clustering method.

We describe several classes of empirical evaluations of methods for network
community detection to demonstrate the artifactual properties and systematic
biases of various community detection objective functions and approximation algorithms. We also
discuss several meta-issues related to community detection algorithms in very
large graphs, including whether or not existing algorithms are sufficiently
powerful to recover interesting communities and whether or not meaningful communities exist at all.
Also in contrast to previous attempts to evaluate community
detection algorithms and/or objective functions,
we consider a size-resolved version of the typical optimization
problem. That is, rather than simply fixing an objective and asking for an
approximation to the best cluster of any size or some fixed partitioning,
we ask for an approximation to the best cluster for every possible size.
This provides a \emph{much} finer lens with which to examine
community detection algorithms, since objective functions and approximation
algorithms often have non-obvious size-dependent behavior.

The rest of the paper is organized as follows. Section~\ref{sec:related} gives
the background and surveys the rich related work in the area of network
community detection.
Then, in Section~\ref{sec:pieces}, we compare structural properties of clusters
extracted by two clustering methods based on two completely different
computational paradigms---a spectral-based graph partitioning
method Local Spectral and a flow-based partitioning algorithm Metis+MQI; and in
Section~\ref{sec:otheralgs}, we extend the analyses by considering related 
heuristic-based clustering algorithms that in practice perform very well.
Section~\ref{sec:scores} then focuses on 11 different objective functions
that attempt to capture the notion of a community as a set of nodes
with better intra- than inter-connectivity.
To understand the performance of various community
detection algorithms at different size scales we compute theoretical lower
bounds on the conductance community-quality score in Section~\ref{sec:bounds}. 
We conclude in Section~\ref{sec:conclusion} with some general observations.

\section{Related Work and Background}
\label{sec:related}
Here we survey related work and summarize our previous work, with an
emphasis on technical issues that motivate this paper.

\subsection{Related work}

A great deal of work has been devoted to finding communities in large
networks, and much of this has been devoted to formalizing the intuition that
a community is a set of nodes that has more and/or better links between its
members than with the remainder of the network. Very relevant to our work is
that of Kannan, Vempala, and Vetta~\cite{kannan04_gbs}, who analyze spectral
algorithms and describe a community concept in terms of a bicriterion
depending on the conductance of the communities and the relative weight of
between-community edges. Flake, Tarjan, and
Tsioutsiouliklis~\cite{FTT03_graph} introduce a similar bicriterion that is
based on network flow ideas, and Flake {\em et al.}~\cite{flake00_efficient}
defined a community as a set of nodes that has more edges pointing inside the
community than to the rest of the network. Similar edge-counting ideas were
used by Radicchi {\em et al.}~\cite{RCCLP04_PNAS} to define and apply the
notions of a strong community and a weak community.

Within the ``complex networks'' community, Girvan and
Newman~\cite{newman02community} proposed an algorithm that used ``betweenness
centrality'' to find community boundaries. Following this, Newman and
Girvan~\cite{newman04community} introduced \emph{modularity} as an \emph{a
posteriori} measure of the overall quality of a graph partition. Modularity
measures internal (and not external) connectivity, but it does so with
reference to a randomized null model. Modularity has been very influential in
recent community detection
literature, and one can use
spectral techniques to approximate
it~\cite{WS05_spectralSDM,newman2006_ModularityPNAS}. However, Guimer\`{a},
Sales-Pardo, and Amaral~\cite{guimera04_fluctuations} and Fortunato and
Barth\'{e}lemy~\cite{Fortunato07_ResolutionPNAS} showed that random graphs
have high-modularity subsets and that there exists a size scale below which
modularity cannot identify communities.

Finally, we should note several other lines of related work. First, the Local
Spectral Algorithm of Andersen, Chung, and Lang~\cite{andersen06local} was
used by Andersen and Lang~\cite{andersen06seed} to find (in a scalable
manner) medium-sized communities in very large social graphs. Second, other
recent work has also focused on developing local and/or near-linear time
heuristics for community detection
include~\cite{clauset05_local}. Third, there also
exists work which views communities from a somewhat different perspective.
For
recent reviews of the large body of work in this area,
see~\cite{gaertler05_clustering,Schaeffer07_survey,For09_TR,LF09_TR}.

\subsection{Background and motivation}
\label{sec:related-background}

\begin{figure*}
\begin{center}
   \includegraphics[width=0.2\textwidth]{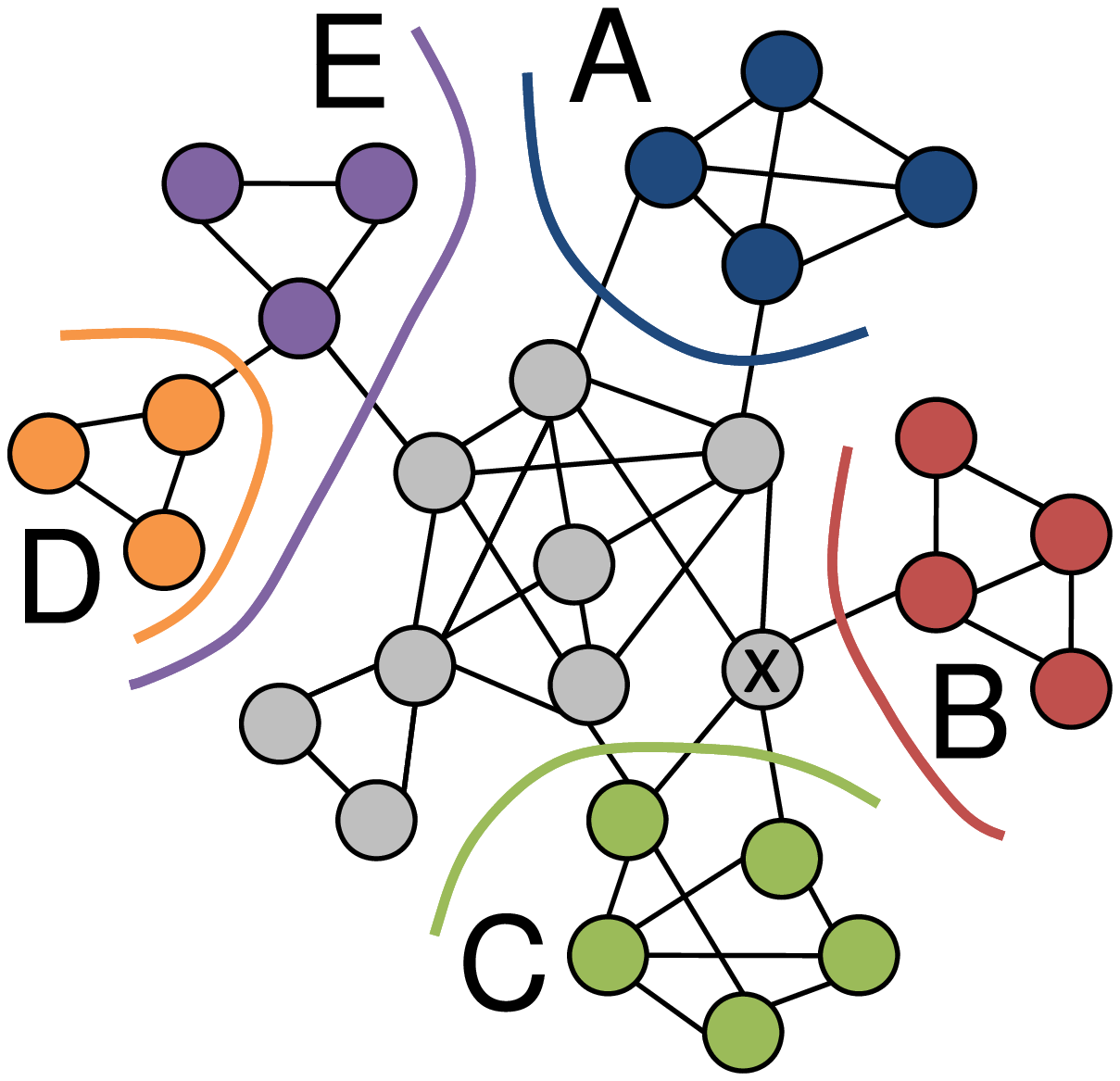} \hspace{5mm}
   \includegraphics[width=0.3\textwidth]{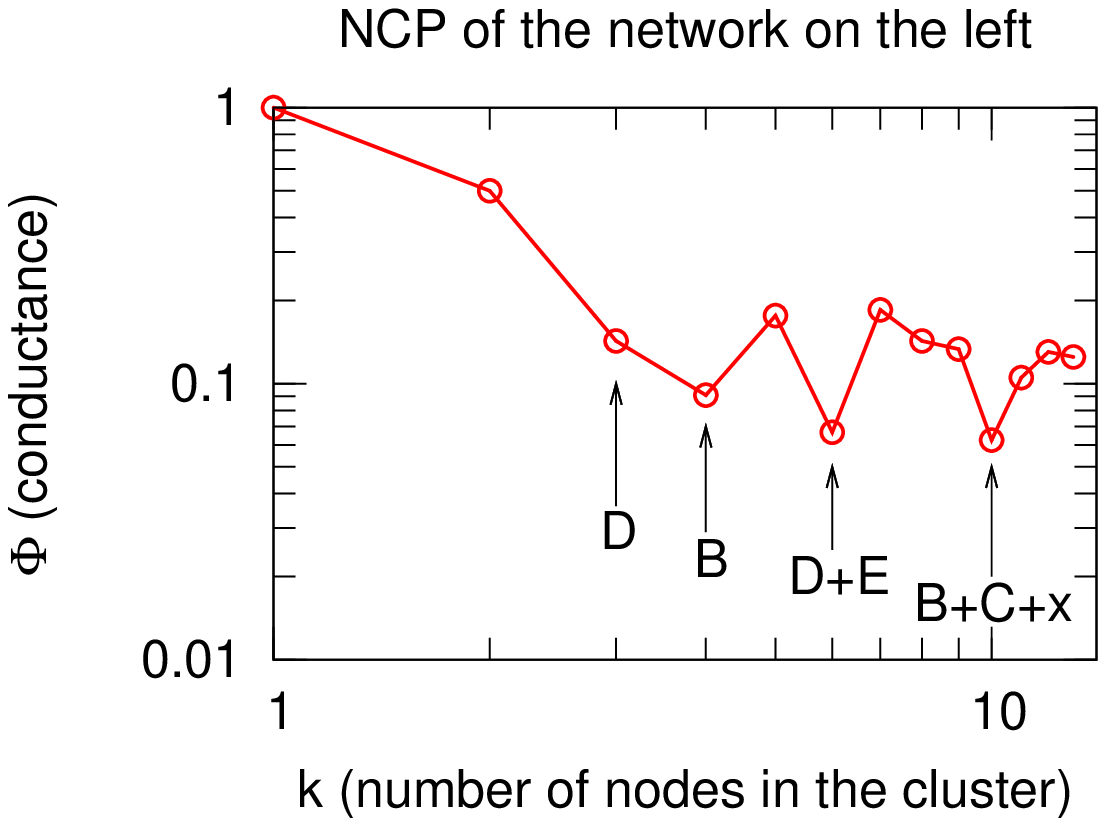}  \hspace{5mm}
   \includegraphics[width=0.3\textwidth]{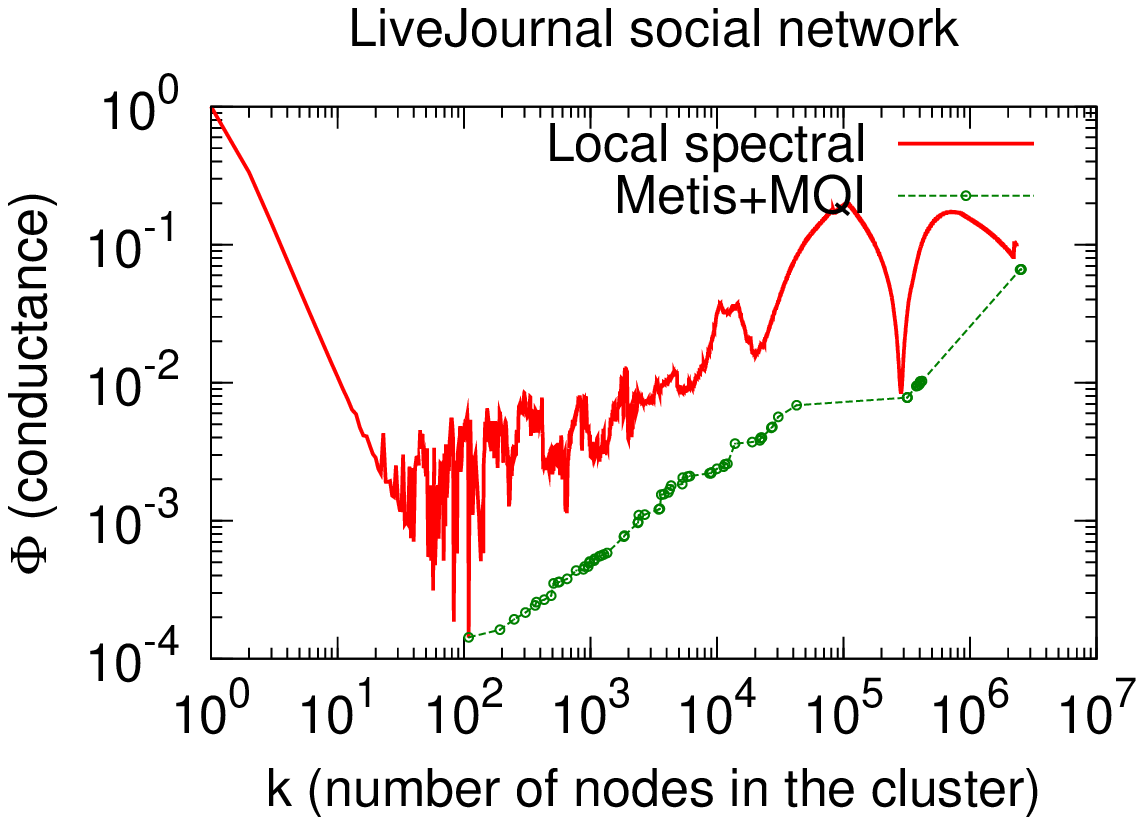}
\end{center}
\vspace{-7mm} \caption{NCP plot (middle) of a small network (left). NCP of
LiveJournal network computed using two different methods.} \vspace{-3mm}
\label{fig:intro}
\end{figure*}

We model each network by an undirected graph, in which nodes represent
entities and edges represent interactions between pairs of entities. We perform
the evaluation of community detection algorithms in a large corpus of over
$100$ social and information networks\footnote{Networks used in this paper
are available at the supporting website~\cite{snap}:
\url{http://snap.stanford.edu/ncp}}. The networks we studied range in size
from tens of nodes and scores of edges up to millions of nodes and tens of
millions of edges; and they were drawn from a wide range of domains, including
large social networks, citation networks, collaboration networks, web graphs,
communication networks, citation networks, internet networks, affiliation
networks, and product co-purchasing networks. In the present work we focus on
a subset of these. In particular, we consider a bipartite authors-to-papers
network of DBLP (\net{AuthToPap-DBLP}), Enron email network
(\net{email-enron}), a co-authorship network of Arxiv Astro physics papers
(\net{CoAuth-astro-ph}), and a social network of \url{Epinions.com}
(\net{Epinions}). See~\cite{snap} for further information and properties of
these networks.

Even though we consider various notions of community score we will primarily
work with {\em conductance}, which arguably is the simplest notion of cluster
quality, as it can be simply thought of as the ratio between the number of edges
inside the cluster and the number of edge leaving the
cluster~\cite{ShiMalik00_NCut,kannan04_gbs}. More formally,
\emph{conductance} $\phi(S)$ of a set of nodes $S$ is
$\phi(S)=c_S/\min(\mbox{Vol}(S), \mbox{Vol}(V \setminus S))$, where $c_S$
denotes the size of the edge boundary, $c_S = |\{(u,v):u \in S,v \notin
S\}|$, and $\mbox{Vol}(S)=\sum_{u\in S} d(u)$, where $d(u)$ is the degree of
node $u$. Thus, in particular, more community-like sets of nodes have {\em
lower} conductance. For example in Figure~\ref{fig:intro}(left), sets $A$ and
$B$ have conductance $\phi(A)=\frac{2}{14} > \phi(B)=\frac{1}{11}$, so the
set of nodes $B$ is more community-like than the set $A$. Conductance
captures a notion of ``surface area-to-volume,'' and thus it is widely-used
to capture quantitatively the gestalt notion of a good network community as a
set of nodes that has better internal- than
external-connectivity~\cite{gaertler05_clustering,Schaeffer07_survey}.

We then generalize the notion of the quality of a single cluster into a size
resolved version. Using a particular measure of network community quality
$f(S)$,
\emph{e.g.}, conductance or one of the other measures described in Section~\ref{sec:scores},
 we then define the \emph{network community profile
(NCP)}~\cite{LLDM08_communities_CONF,LLDM08_communities_TR} that
characterizes the quality of network communities as a function of their size.
For every $k$ between $1$ and half the number of nodes in the
network\footnote{Note that one only needs to consider clusters of sizes up to
half the number of nodes in the network since $\phi(S)=\phi(V\setminus S)$.},
we define $\Phi(k) = \min_{|S|=k}f(S)$. That is, for every possible community
size $k$, $f(k)$ measures the score of the most community-like set of nodes of that size,
and the NCP measures $\Phi(k)$ as a function of $k$.

For example, in Figure~\ref{fig:intro}(middle) we use conductance as a
measure of cluster quality and for $k=4$, among all sets of $4$-nodes, $B$
has best conductance, and thus $\Phi(4)=\frac{1}{11}$. Similarly, $D$ and
$D$+$E$ denote the best conductance sets on $3$ and $6$ nodes, respectively.

Just as the magnitude of the conductance provides information about how
community-like is a set of nodes, the shape of the NCP provides insight into
how well expressed are network communities as a function of their size.
Moreover, the NCP also provides a lens to examine the quality of clusters of
various sizes. Thus in the majority of our experiments we will examine and
compare different clustering algorithms and objective functions through
various notions of the NCP plot and other kinds of structural metrics of
clusters and how they depend/scale with the size of the cluster.

Moreover, the shape of the NCP is also interesting for a very different reason.
It gives us a powerful way to quantify and summarize the large-scale
community structure of networks.
We~\cite{LLDM08_communities_CONF,LLDM08_communities_TR} found that the NCP
behaves in a characteristic manner for a range of large social and
information networks: when plotted on log-log scales, the NCP tends to have a
universal ``V'' shape (Figure~\ref{fig:intro}(right)). Up to a size scale of
about $100$ nodes, the NCP decreases, which means that the best-possible clusters
are getting progressively better with the increasing size. The NCP then reaches
the minimum at around $k=100$ and then gradually increases again, which means
that at larger size scales network communities become less and less
community-like. (This should be contrasted with behavior for mesh-like
networks, road networks, common network generation models, and small
commonly-studied networks, for which the NCP is either flat or
downward-sloping~\cite{LLDM08_communities_CONF,LLDM08_communities_TR}.) The
shape of the NCP can be explained by an onion-like ``nested core-periphery''
structure, where the network consists of a large 
core (slightly denser and more expander-like than the full graph, but which itself has a core-periphery structure)
and a large number of small very well-connected
communities barely connected to the
core~\cite{LLDM08_communities_CONF,LLDM08_communities_TR}.  In this context,
it is important to understand the characteristics of
various community detection algorithms in order to make sure that the shape
of NCP is a property of the network rather than an artifact of the approximation
algorithm or the function that formalizes the notion of a network community.

\section{Comparison of algorithms}
\label{sec:algs}

We compare different clustering algorithms and heuristics. We focus our
analyses on two aspects. First, we are interested in the quality of the clusters that
various methods are able to find. Basically, we would like to understand how
well algorithms perform in terms of optimizing the notion of community
quality (conductance in this case). Second, we are interested in quantifying
the structural properties of the clusters identified by the algorithms. As we
will see, there are fundamental tradeoffs in network community detection---for
a given objective function, approximation algorithms are often biased in a sense that they consistently find clusters
with particular internal structure.

We break the experiments into two parts. First, we compare two graph
partitioning algorithms that are theoretically well understood and are based
on two very different approaches: a spectral-based Local Spectral partitioning
algorithm, and the flow-based Metis+MQI. Then we consider several
heuristic approaches to network community detection that work well in
practice.

\subsection{Flow and spectral methods}
\label{sec:pieces}
In this section we compare the Local Spectral Partitioning
algorithm~\cite{andersen06local} with the flow-based Metis+MQI algorithm.
The latter is a surprisingly effective heuristic method for finding
low-conductance cuts, which consists of first using the fast graph
bi-partitioning program Metis~\cite{karypis98_metis} to split the
graph into two equal-sized pieces, and then running MQI, an exact 
flow-based technique~\cite{Gallo:1989,kevin04mqi} for finding the 
lowest conductance cut whose small side in contained in one of the
two half-graphs chosen by Metis.

\begin{figure*}[t!]
\begin{center}
\includegraphics[width=0.32\linewidth]{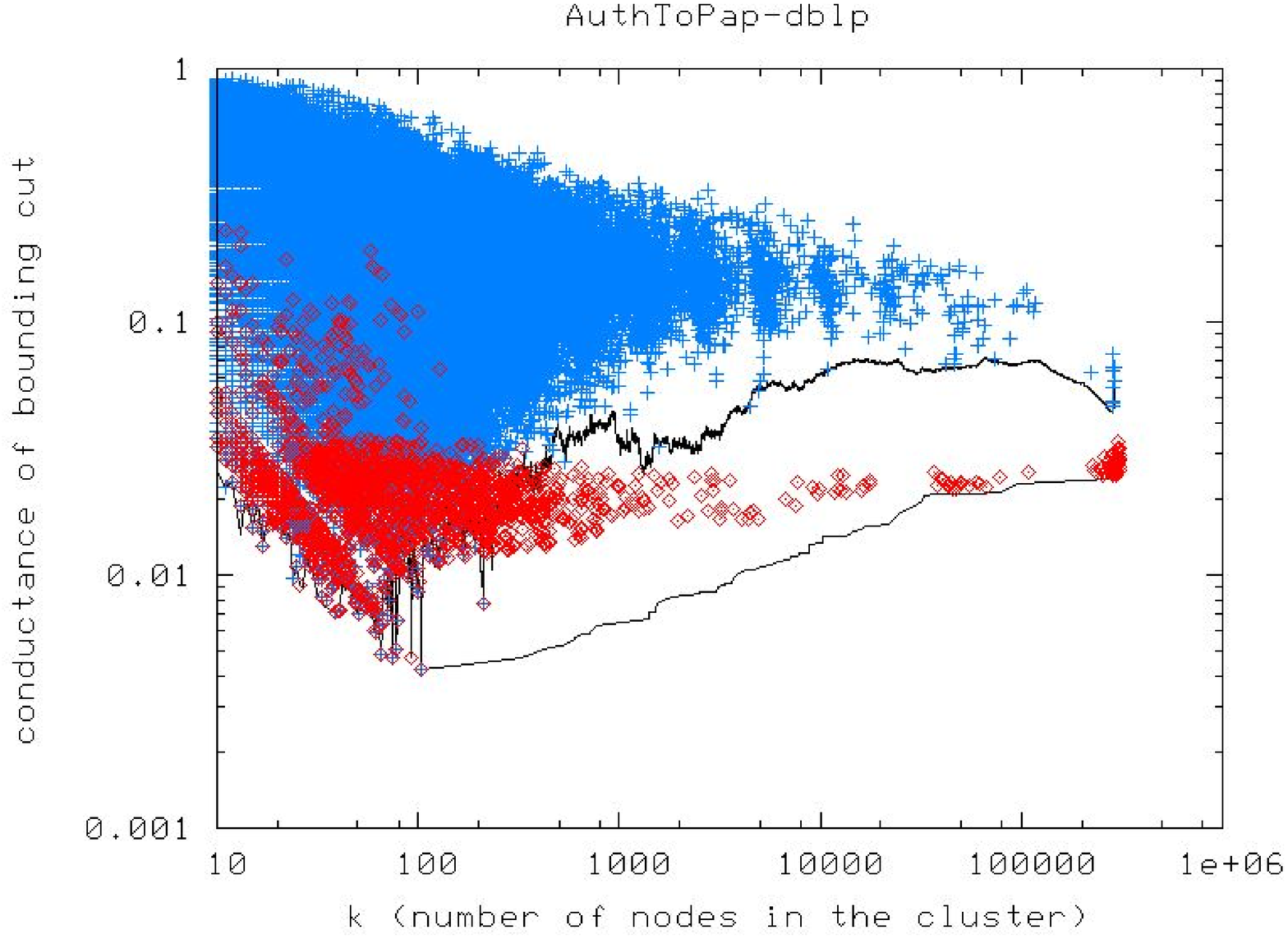}
\includegraphics[width=0.32\linewidth]{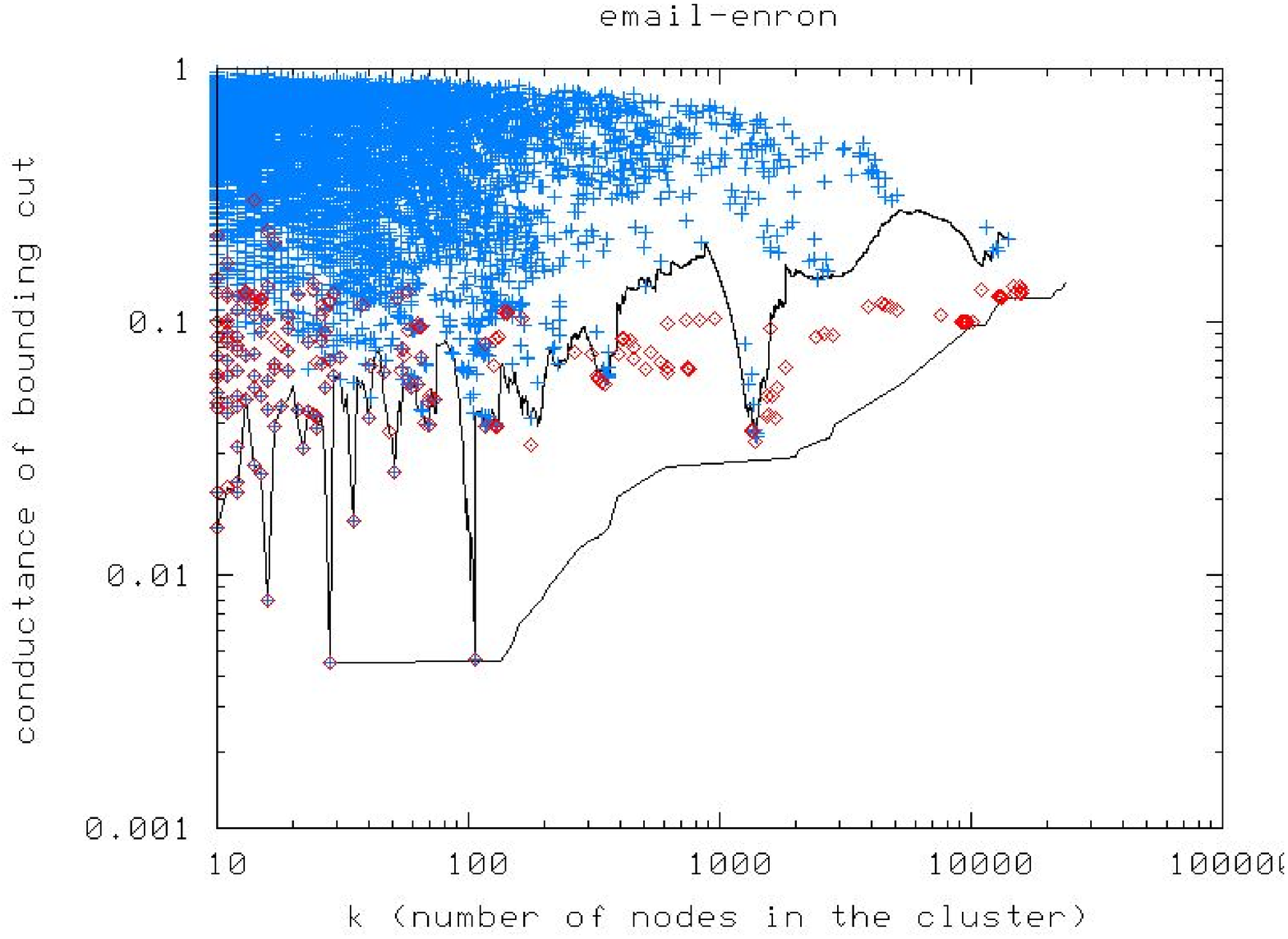}
\includegraphics[width=0.32\linewidth]{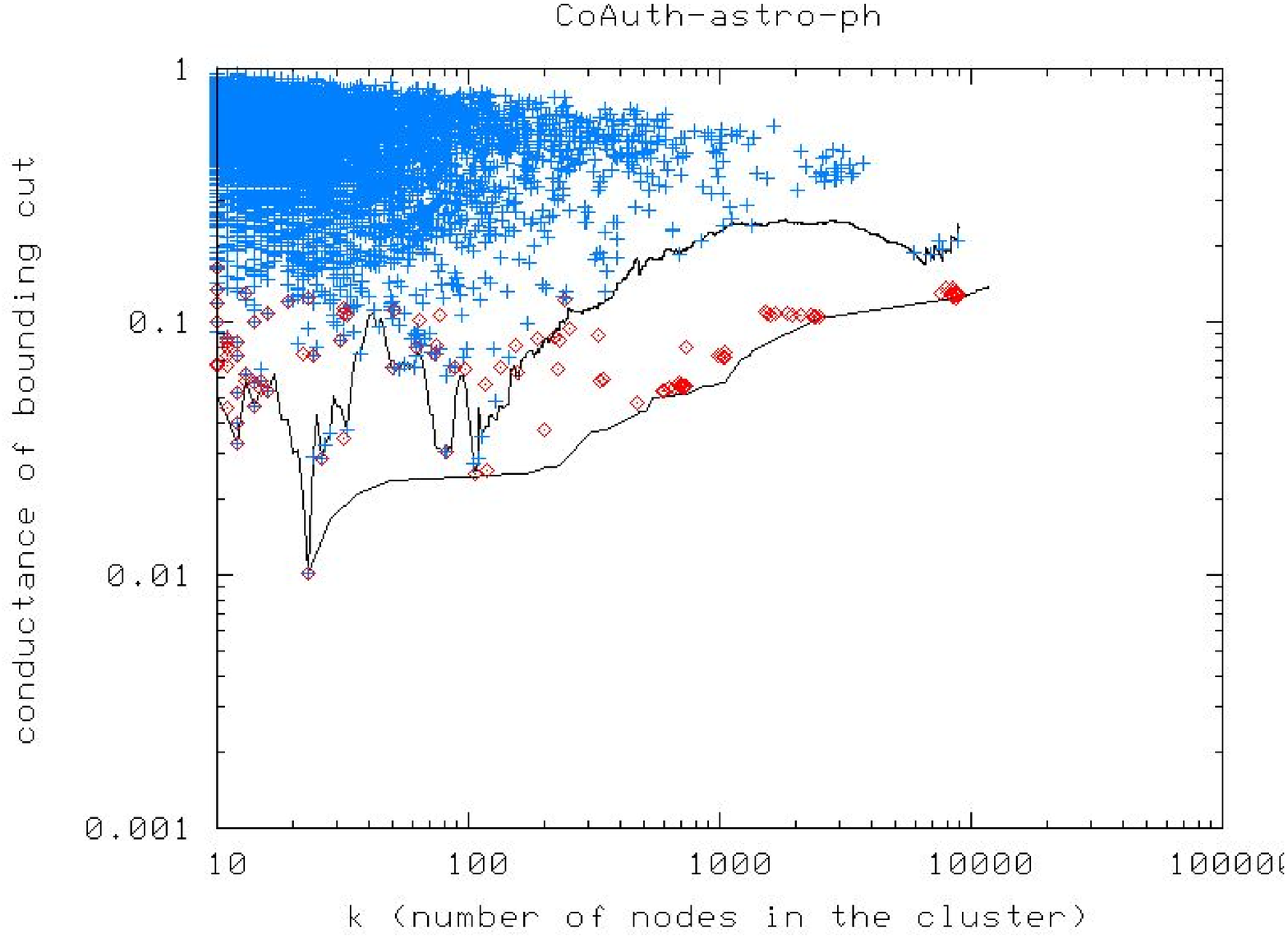}\\
Conductance of connected clusters found by Local Spectral (blue) and Metis+MQI (red)\\
\includegraphics[width=0.32\linewidth]{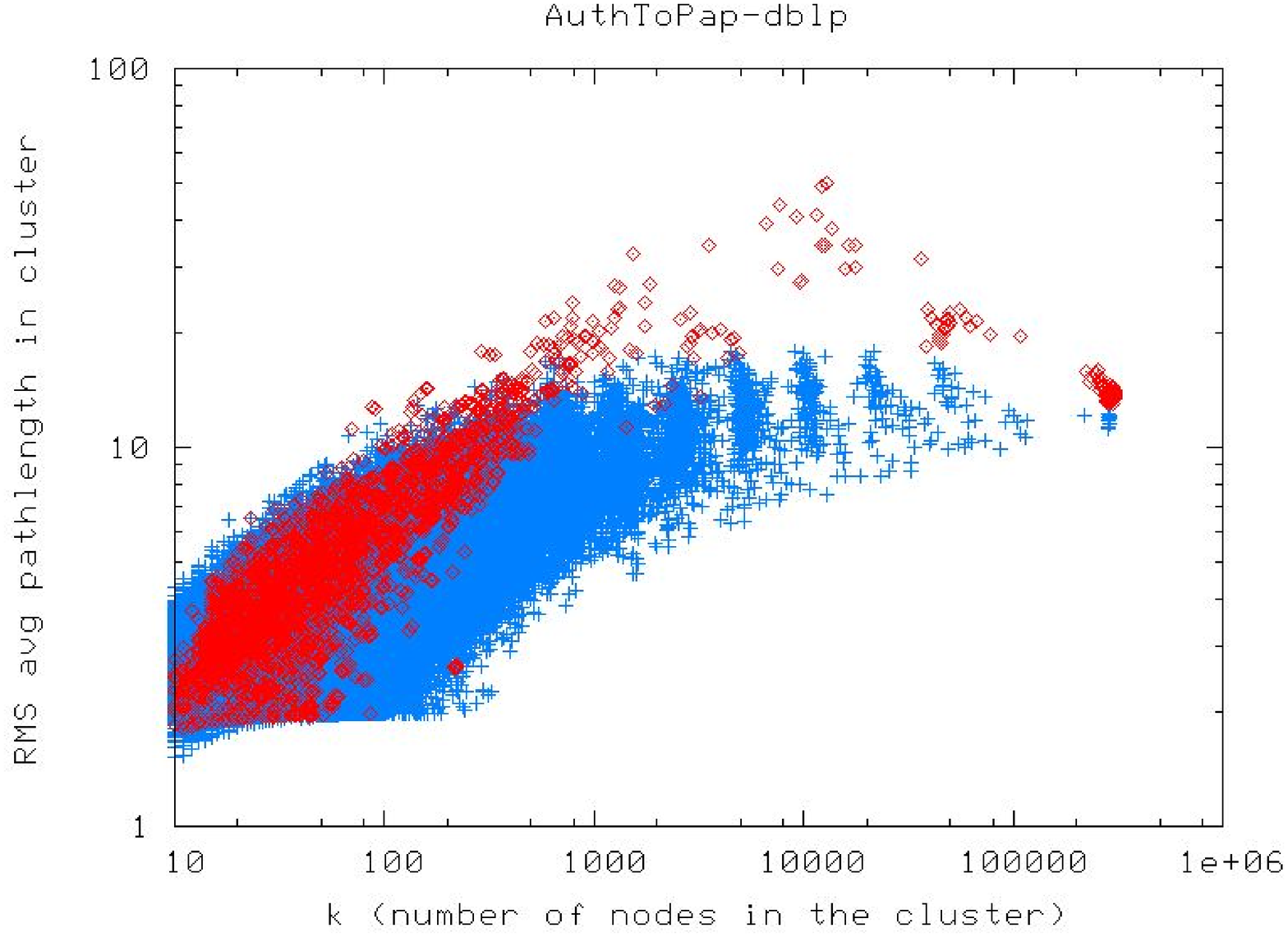}
\includegraphics[width=0.32\linewidth]{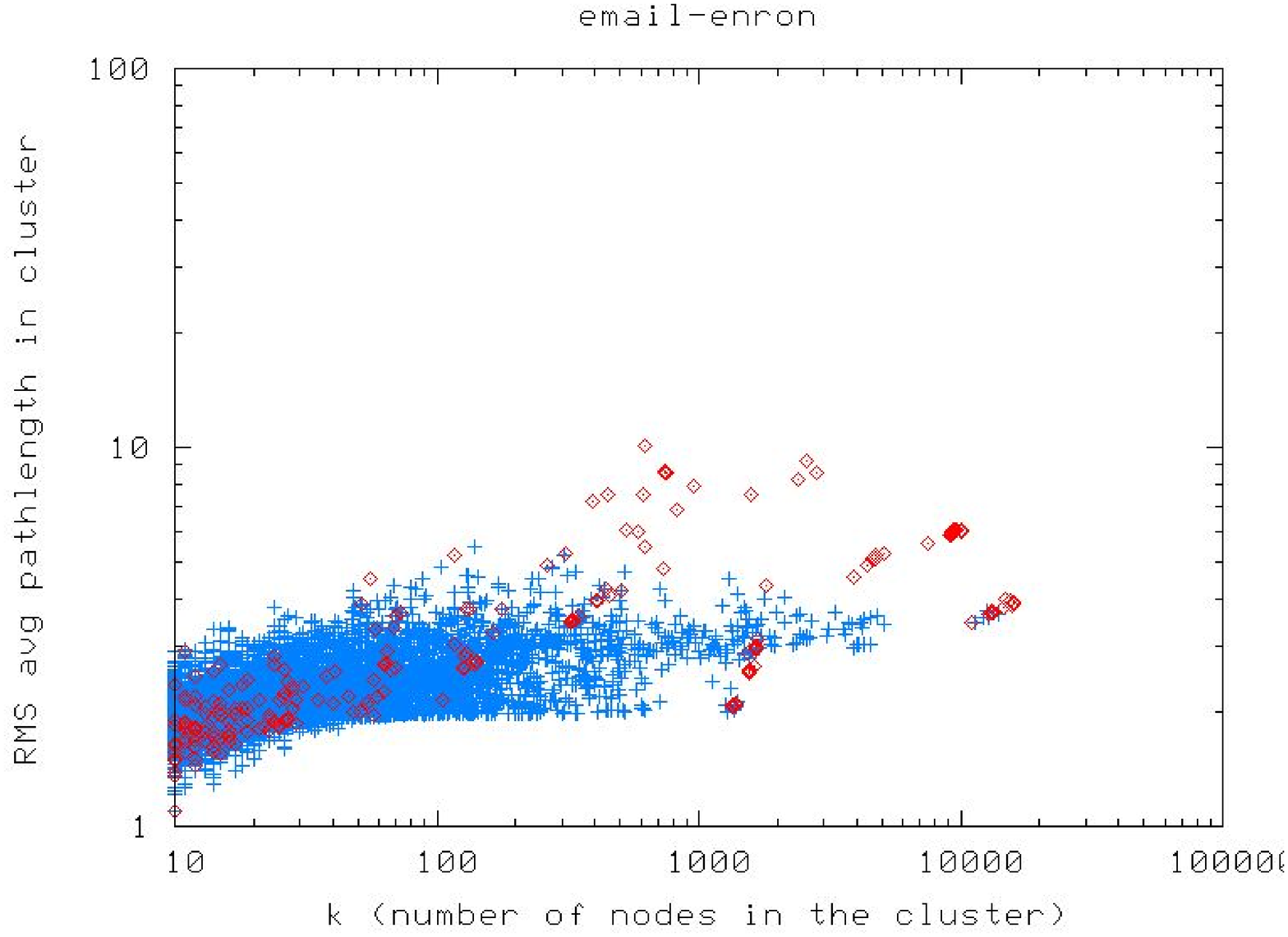}
\includegraphics[width=0.32\linewidth]{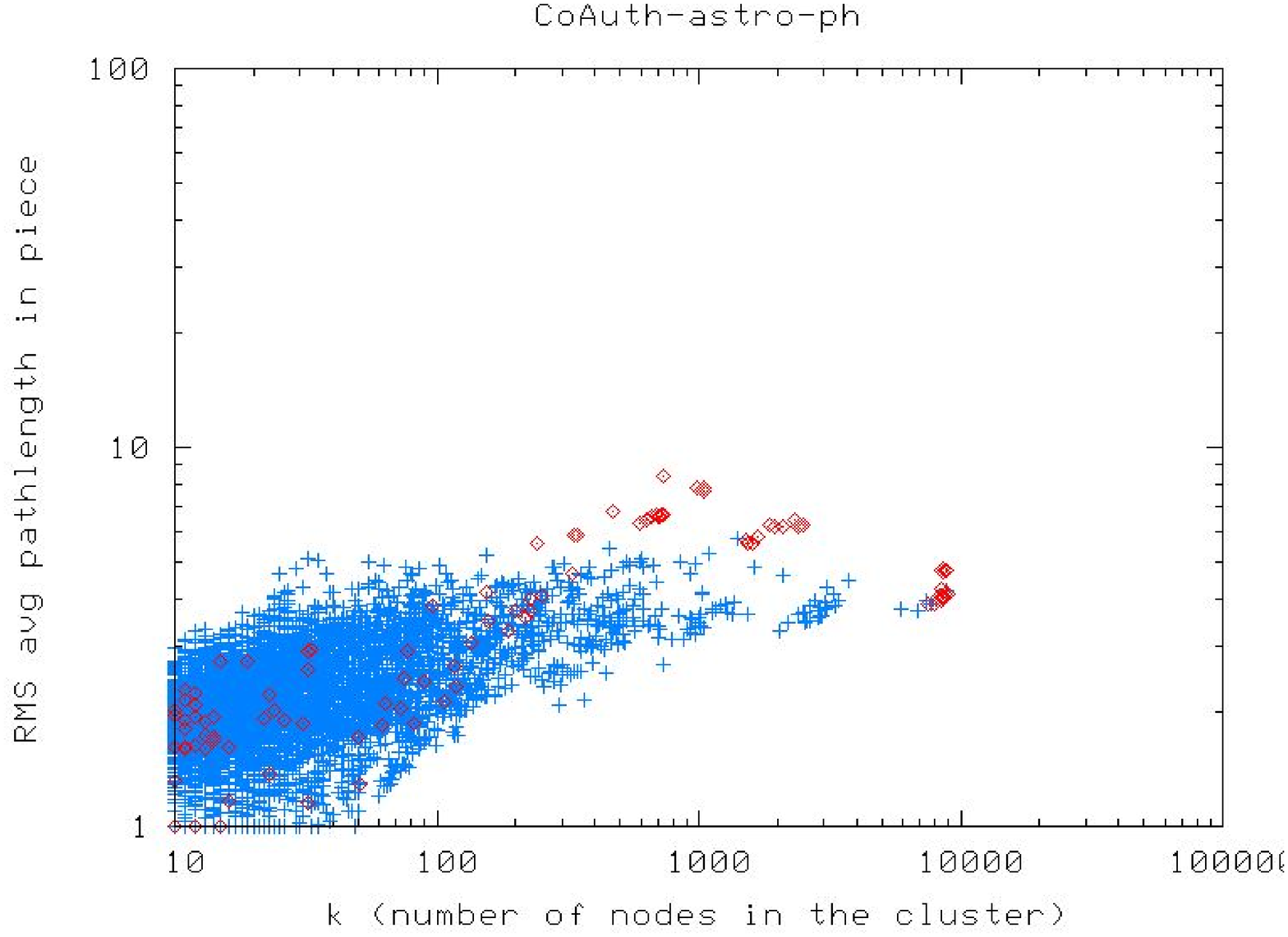}\\
Cluster compactness: average shortest path length\\
\includegraphics[width=0.32\linewidth]{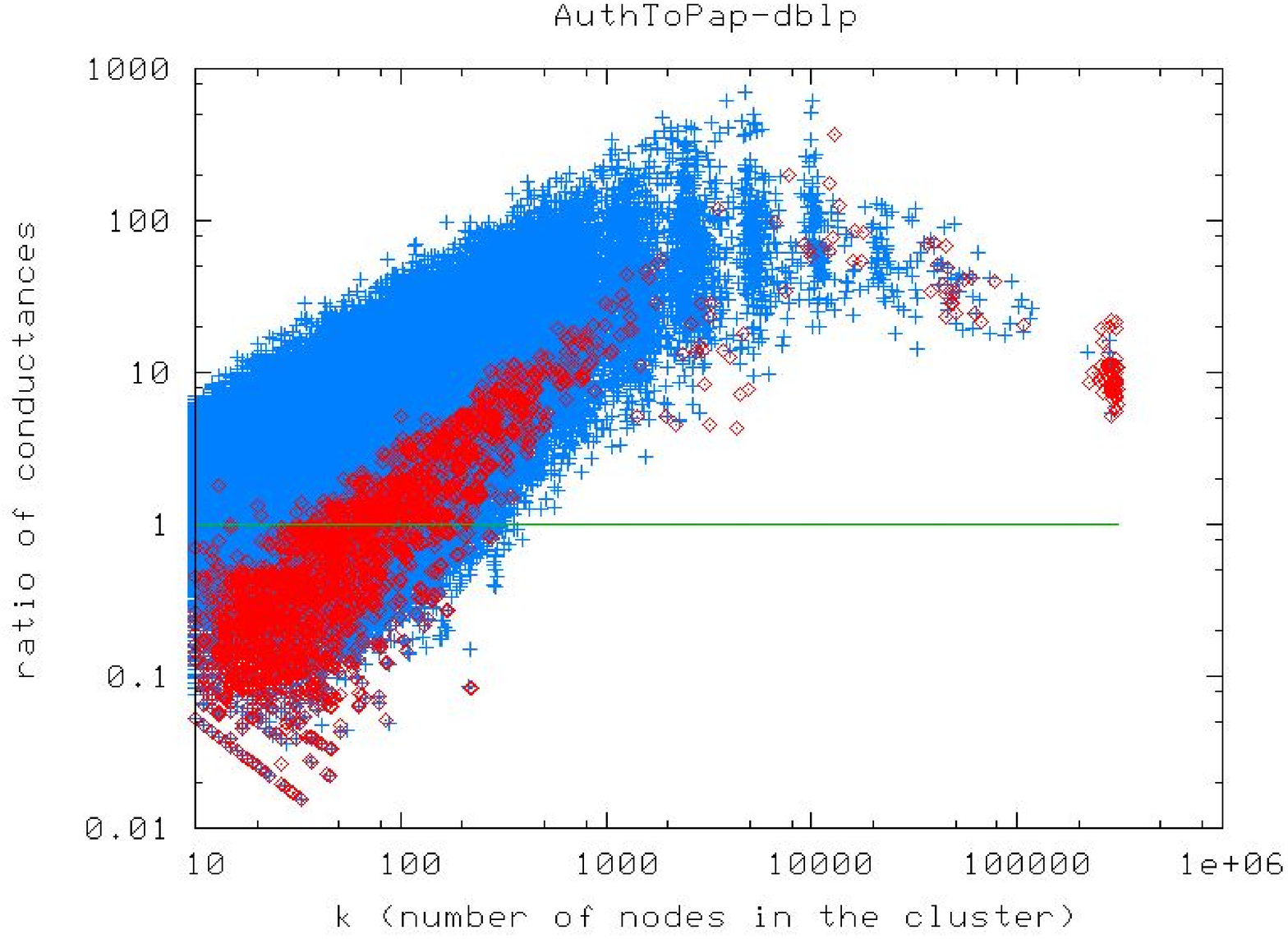}
\includegraphics[width=0.32\linewidth]{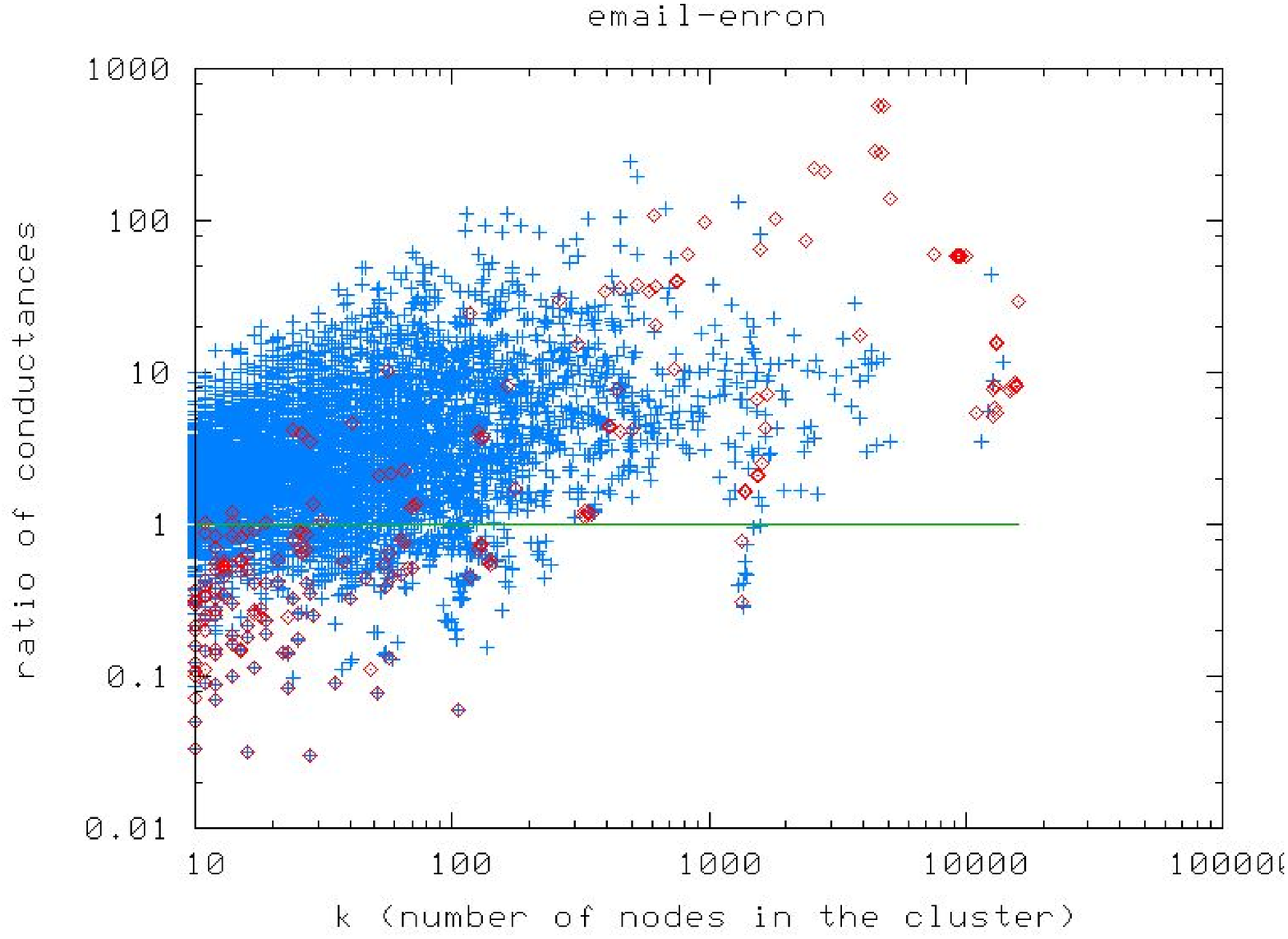}
\includegraphics[width=0.32\linewidth]{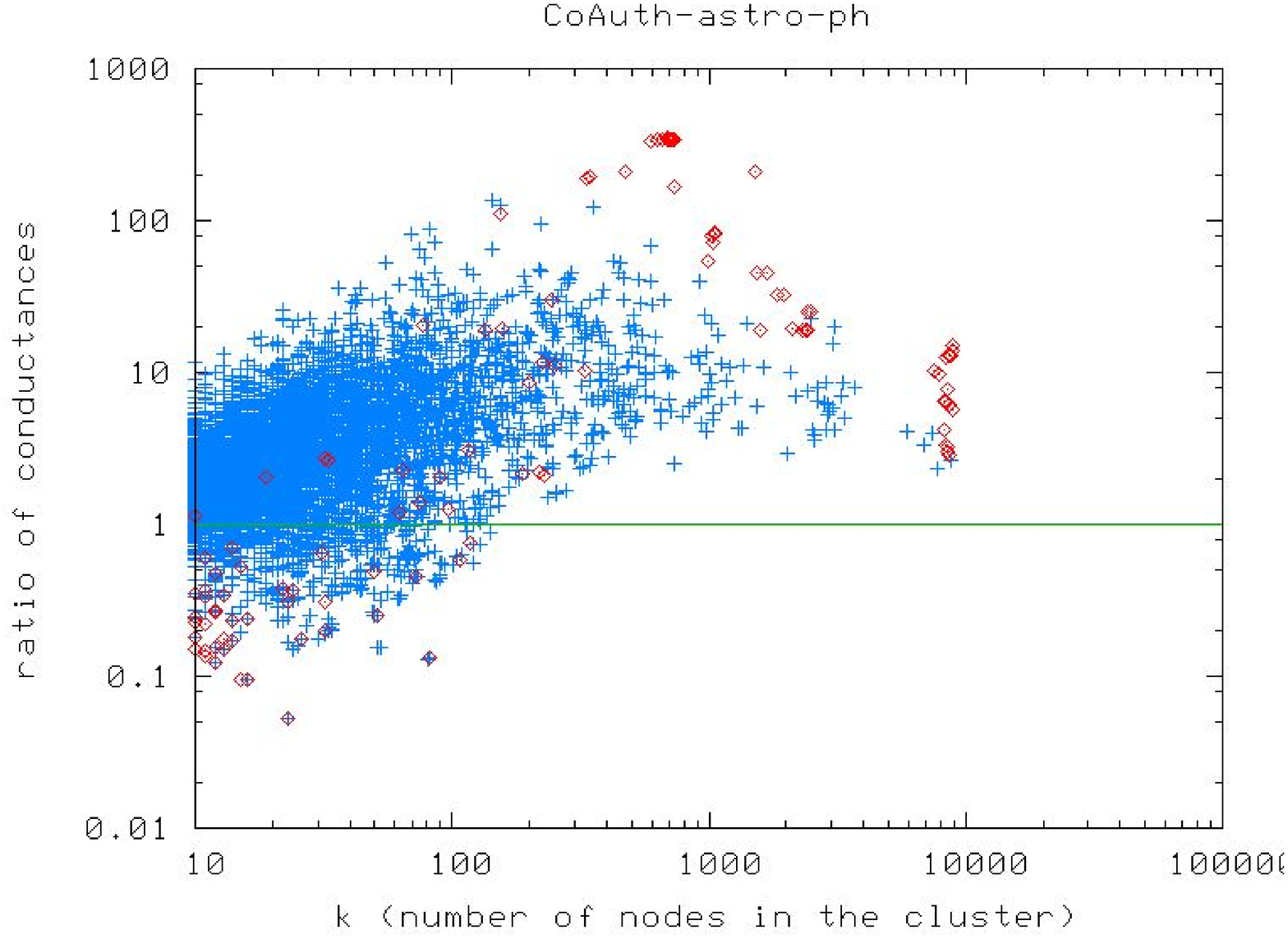}\\
Cluster compactness: external vs. internal conductance
\end{center}
\vspace{-5mm}
\caption{
Comparison of Local Spectral (blue) and Metis+MQI (red) on connected clusters.
Top: the conductance of the bounding cut.
Middle: the average shortest path length in the cluster.
Bottom: the ratio of the external conductance to the
internal conductance.
Generally Metis+MQI yields better cuts
while Local Spectral yields clusters that are more
compact: they have shorter path lengths and better internal connectivity.
}
\vspace{-3mm}
\label{compactness-vs-cuts-fig}
\end{figure*}

Each of those two methods (Local Spectral and Metis+MQI) was run
repeatedly with randomization on each of our graphs, to produce a large
collection of candidate clusters of various sizes, plus a
lower-envelope curve. The lower-envelope curves for the two algorithms
were the basis for the plotted NCP's in the earlier
paper~\cite{LLDM08_communities_CONF}. 
In the current paper the lower-envelope curves for
Local Spectral and Metis+MQI are plotted respectively as a red line and a 
green line in Figure~\ref{fig:intro}(right), and as pairs of black lines in
Figure~\ref{compactness-vs-cuts-fig}(top) and Figures~\ref{other-algos-fig} 
and~\ref{other-algos-fig-LBfig}. 
Note that the Metis+MQI curves are generally lower,
indicating that this method is generally better than Local Spectral
at the nominal task of finding cuts with low conductance.

However, as we will demonstrate using the scatter plots of 
Figure~\ref{compactness-vs-cuts-fig}, the clusters found by the Local
Spectral Method often have other virtues that compensate for their
worse conductance scores. 
As an extreme example, many of
the raw Metis+MQI clusters are internally disconnected, which seems
like a very bad property for an alleged community. By contrast, the
Local Spectral Method always returns connected clusters. 
Acknowledging that this is a big advantage for Local Spectral,
we then modified the collections of raw Metis+MQI clusters by
splitting every internally disconnected cluster into its various
connected components. Then, in all scatter plots of Figure~\ref{compactness-vs-cuts-fig}, 
blue dots represent raw Local Spectral clusters, which are internally connected,
while red dots represent broken-up Metis+MQI clusters, which are also internally connected.

Let us now consider the top row of scatter plots of
Figure~\ref{compactness-vs-cuts-fig} which compares the conductance
scores (as a function of cluster size) of the collections of clusters
produced by the two algorithms. The cloud of blue points (Local
Spectral clusters) lies generally above the cloud of red points
(Metis+MQI clusters), again illustrating that Local Spectral tends to be a
weaker method for minimizing conductance score.  In more detail, we
find that Local Spectral and Metis+MQI tend to identify similar
pieces at very small scales, but at slightly larger scales a gap opens
up between the red cloud and the blue cloud. At those intermediate
size scales, Metis+MQI is finding lower conductance cuts than Local
Spectral.

However, the Local Spectral algorithm returns pieces that are internally more
\emph{compact}. This is shown in the middle row of
Figure~\ref{compactness-vs-cuts-fig} where for each of the (connected) pieces
for which we plotted a conductance in the top row, we are now plotting the
average shortest path length between random node pairs in that piece. In
these plots, we see that in the same size range where Metis+MQI is generating
clearly lower conductance connected sets, Local Spectral is generating pieces
with clearly shorter internal paths, \emph{i.e.}, smaller diameter sets. In other words,
the Local Spectral pieces are more ``compact.'' This effect is especially
pronounced in the DBLP affiliation network, while it also shows up in the
Enron email network and the astrophysics collaboration network. Moreover, we
made similar observations also for many other datasets (plots not shown).

Finally, in the bottom row of Figure~\ref{compactness-vs-cuts-fig} we
introduce the topic of internal vs. external cuts, which is something that
none of the existing algorithms is \emph{explicitly}
optimizing. These are again scatter plots showing the same set of Local
Spectral and Metis+MQI pieces as before, but now the $y$-axis is external
conductance divided by internal conductance. External conductance is the
quantity that we usually plot, namely the conductance of the cut which
separates the cluster from the graph. Internal conductance is the score of a
low conductance cut {\em inside} the cluster. That is, we take the induced
subgraph on the cluster's nodes and then find best conductance cut inside the
cluster.

We then compare the ratios of the conductance of the bounding cut and the
internal conductance. Intuitively, good and compact communities should have
small ratios, ideally below 1.0, which would mean that those clusters are well
separated from the rest of the network and that they are also internally well-connected and
hard to cut again. However, the three bottom-row plots of
Figure~\ref{compactness-vs-cuts-fig} show the ratios. Points above the
horizontal line are clusters which are easier to cut internally than they
were to be cut from the rest of the network; while points below the line are
clusters that were relatively easy to cut from the network and are internally
well-connected. Notice that here the distinction between the two methods is
less clear. On the one hand, Local Spectral finds clusters that have worse (higher)
bounding cut conductance, while such clusters are also internally more
compact (have internal cuts of higher conductance). On the other hand,
Metis+MQI finds clusters that have better (lower) bounding cut conductance
but are also internally easy to cut (have internal cut of lower conductance).
Thus when one takes the ratio of the two quantities we observe qualitatively
similar behaviors.  However, notice that Local Spectral seem to return
clusters with higher variance in the ratio of external-to-internal
conductance. At small size scales Metis+MQI tends to give clusters of
slightly better (lower) ratio, while at larger clusters the advantage goes to
Local Spectral. This has interesting consequence for the applications of graph
partitioning since (depending on the particular application domain and the
sizes and properties of clusters one aims to extract) either Local Spectral or
Metis+MQI may be the method of choice.

Also, notice that there are mostly no ratios well below $1.0$, except for very
small sizes. This is important, as it seems to hint that large clusters are
relatively hard to cut from the network, but are then internally easy to split
into multiple sub-clusters. This shows another aspect of our findings: small
communities below $\approx 100$ nodes are internally compact and well
separated from the remainder of the network, whereas larger clusters are so
hard to separate that cutting them from the network is more expensive than
cutting them internally.
Community-like sets of nodes that are better connected internally than externally don't 
seem to exist in large real-world networks, except at very small size scales.

\begin{figure*}[t!]
\begin{center}
\vspace{-1mm}
{\includegraphics[width=0.24\linewidth]{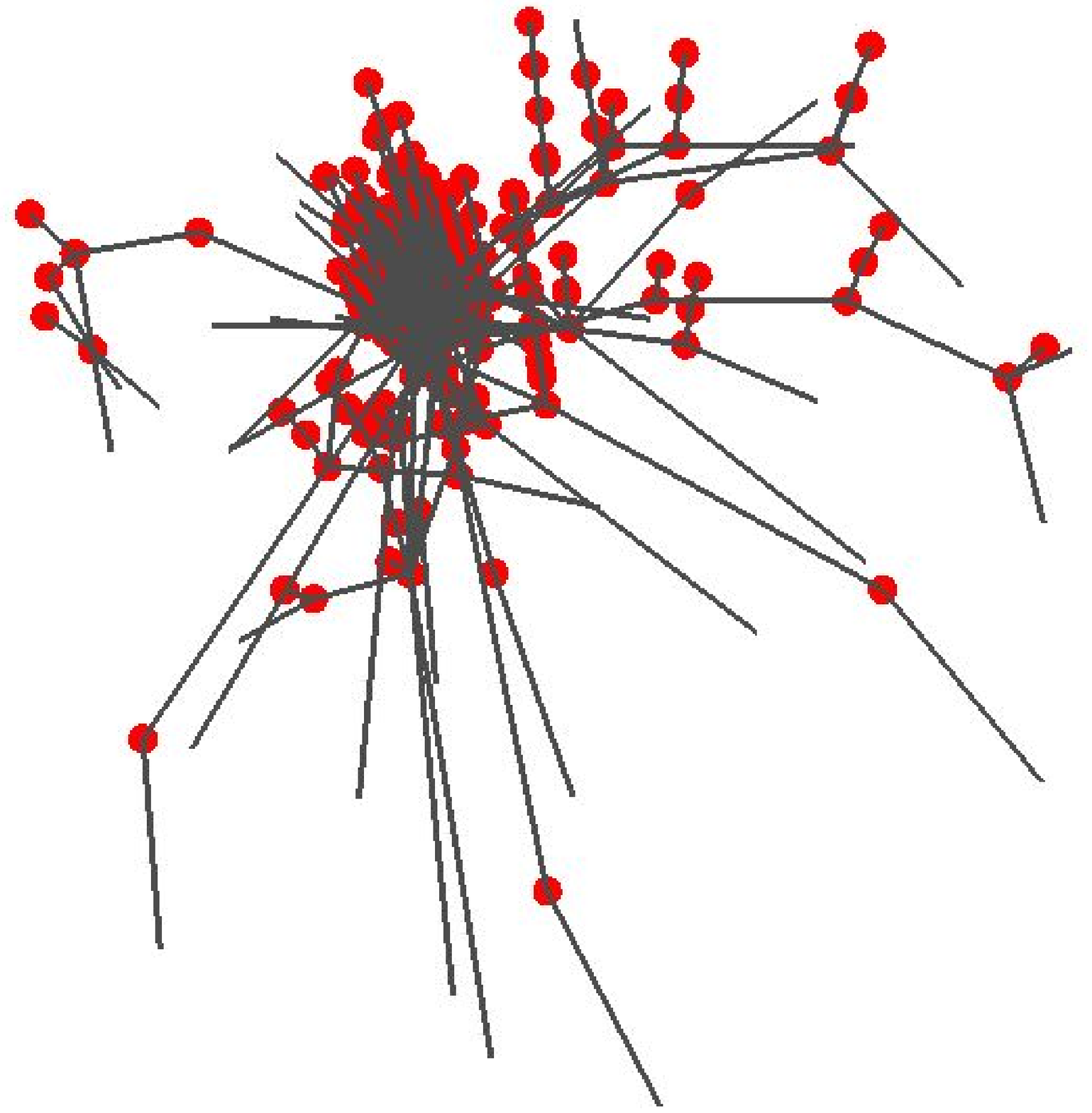}}
{\includegraphics[width=0.24\linewidth]{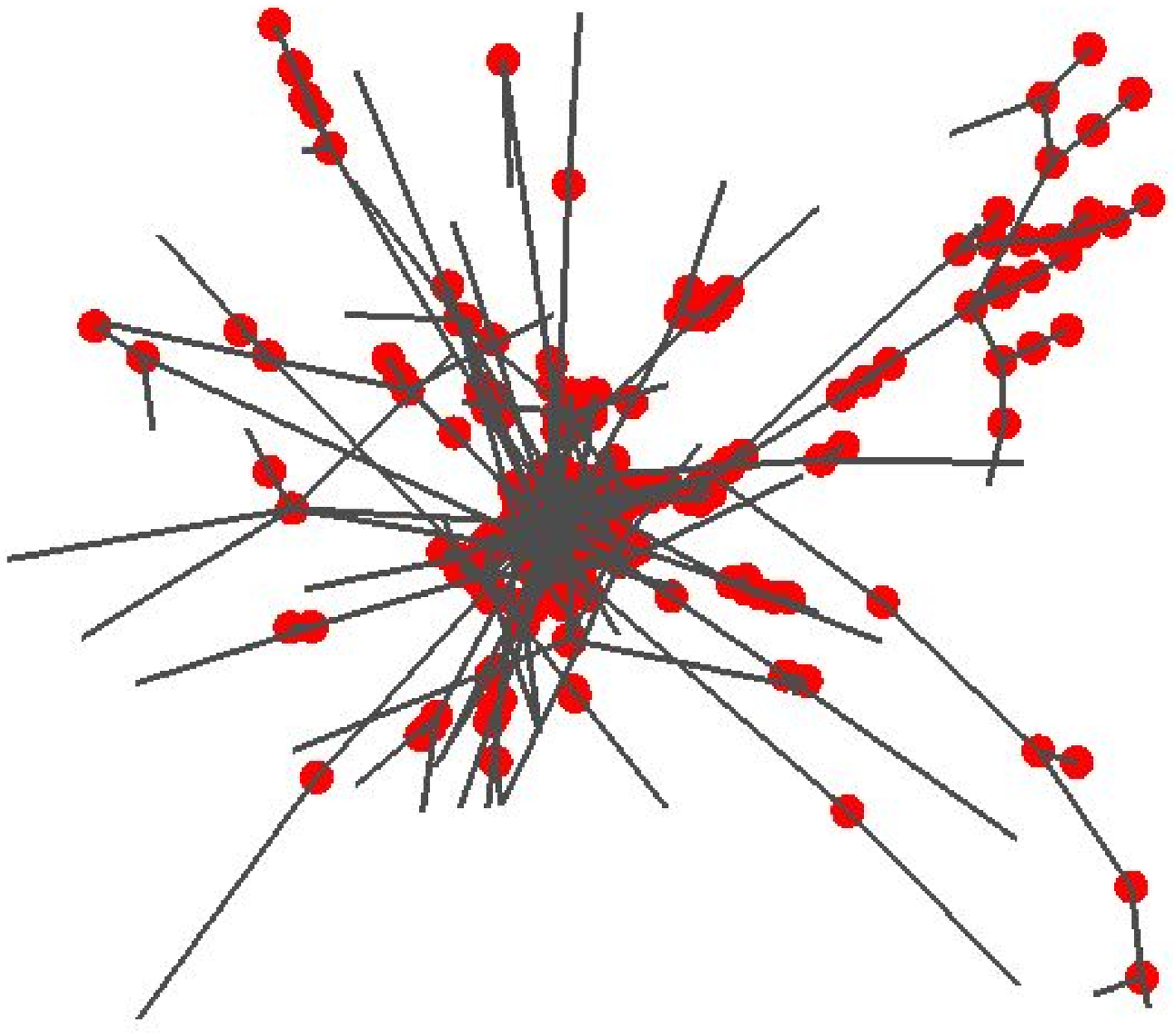}}
{\includegraphics[width=0.22\linewidth]{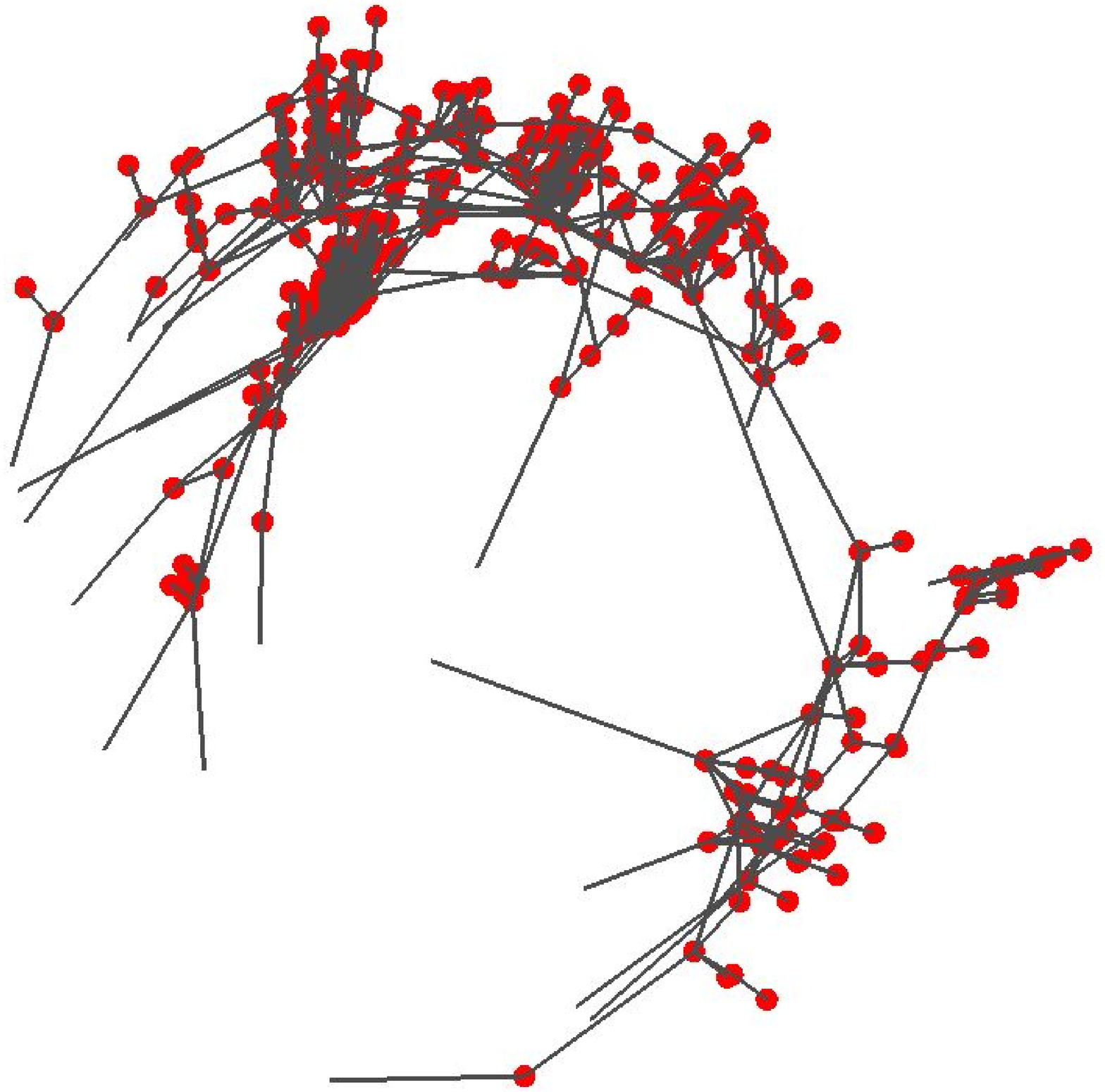}}
{\includegraphics[width=0.22\linewidth]{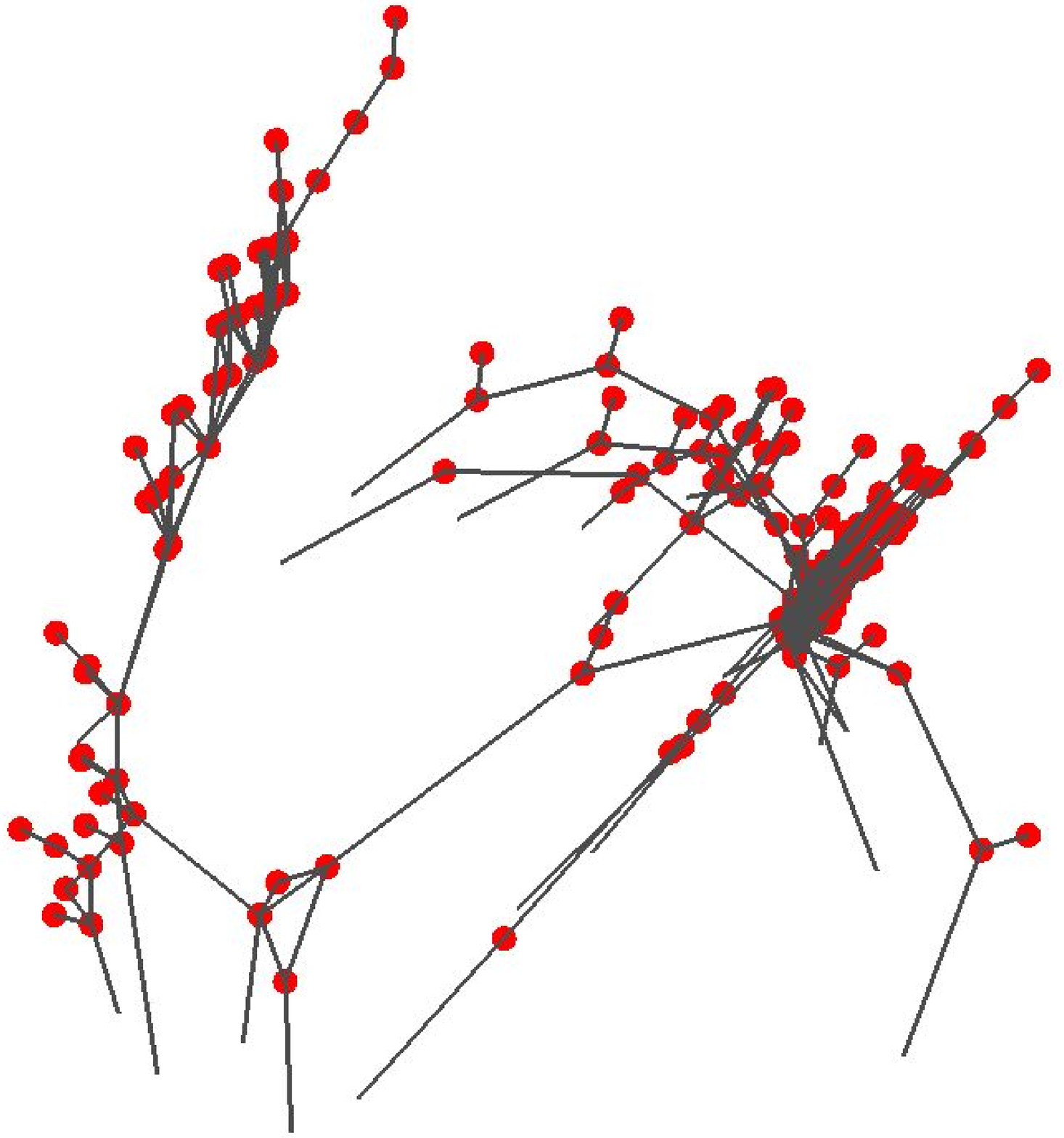}}
\end{center}
\vspace{-10mm} \caption{ Two examples of clusters found by the LocalSpectral
algorithm (on the left) and two from the Metis+MQI algorithm (on the right).
Note that the Local Spectral clusters are more compact---they are tighter and
have smaller diameter since the algorithm has difficulty pushing probability
mass down long extended paths---while the Metis+MQI clusters are more
sprawling---they have larger diameter and more diverse internal structure,
but better conductance scores. In both cases, we have shown communities with
ca. $500$ nodes (many of nodes overlap at resolution of this figure). }
\vspace{-3mm} \label{sprawl-plots}
\end{figure*}

Last, in Figure~\ref{sprawl-plots}, we further illustrate the differences
between spectral and flow-based clusters by drawing some example subgraphs.
The two subgraphs shown on the left of Figure~\ref{sprawl-plots} were found
by Local Spectral, while the two subgraphs shown on the right of
Figure~\ref{sprawl-plots} were found by Metis+MQI.  These two pairs of
subgraphs have a qualitatively different appearance: Metis+MQI
pieces look longer and stringier than the Local Spectral pieces. All of
these subgraphs contain roughly 500 nodes, which is about the size scale
where the differences between the algorithms start to show up. In these
cases, Local Spectral has grown a cluster out a bit past its natural
boundaries (thus the spokes), while Metis+MQI has strung together a couple of
different sparsely connected clusters. (We remark that the tendency of Local
Spectral to trade off cut quality in favor of piece compactness isn't just an
empirical observation, it is a well understood consequence of the theoretical
analysis of spectral partitioning methods.)

\subsection{Other algorithms}
\label{sec:otheralgs}

Next we consider various other, mostly heuristic, algorithms and compare
their  performance in extracting clusters of various sizes. As a point of
reference we use results obtained by the Local Spectral and Metis+MQI
algorithms.

We have extensively experimented with several variants of the global spectral
method, both the usual eigenvector-based embedding on a line, and an
SDP-based embedding on a hypersphere, both with the usual hyperplane-sweep
rounding method and a flow-based rounding method which includes MQI as the
last step.  In addition, special post-processing can be done to obtain either
connected or disconnected sets. 

We also experimented with a
practical version of the Leighton-Rao
algorithm~\cite{Leighton:1988,Leighton:1999}, similar to the implementation
described in~\cite{kevin93_finding,kevin04mqi}. These results are especially
interesting because the Leighton-Rao algorithm, which is based on
multi-commodity flow, provides a completely independent check on Metis, and
on spectral methods generally.
%%% COMBINE TWO PARS
The Leighton-Rao algorithm has two phases. In the first phase, edge
congestions are produced by routing a large number of commodities through the
network. We adapted our program to optimize conductance (rather than ordinary
ratio cut score) by letting the expected demand between a pair of nodes be
proportional to the product of their degrees. In the second phase, a rounding
algorithm is used to convert edge congestions into actual cuts.  Our method
was to sweep over node orderings produced by running Prim's Minimum Spanning
Tree algorithm on the congestion graph, starting from a large number of
different initial nodes, using a range of different scales to avoid quadratic
run time. We used two variations of the method, one that produces connected
sets, and another one that can also produce disconnected sets.

\begin{figure*}[t!]
\begin{center}
    \includegraphics[width=0.32\linewidth]{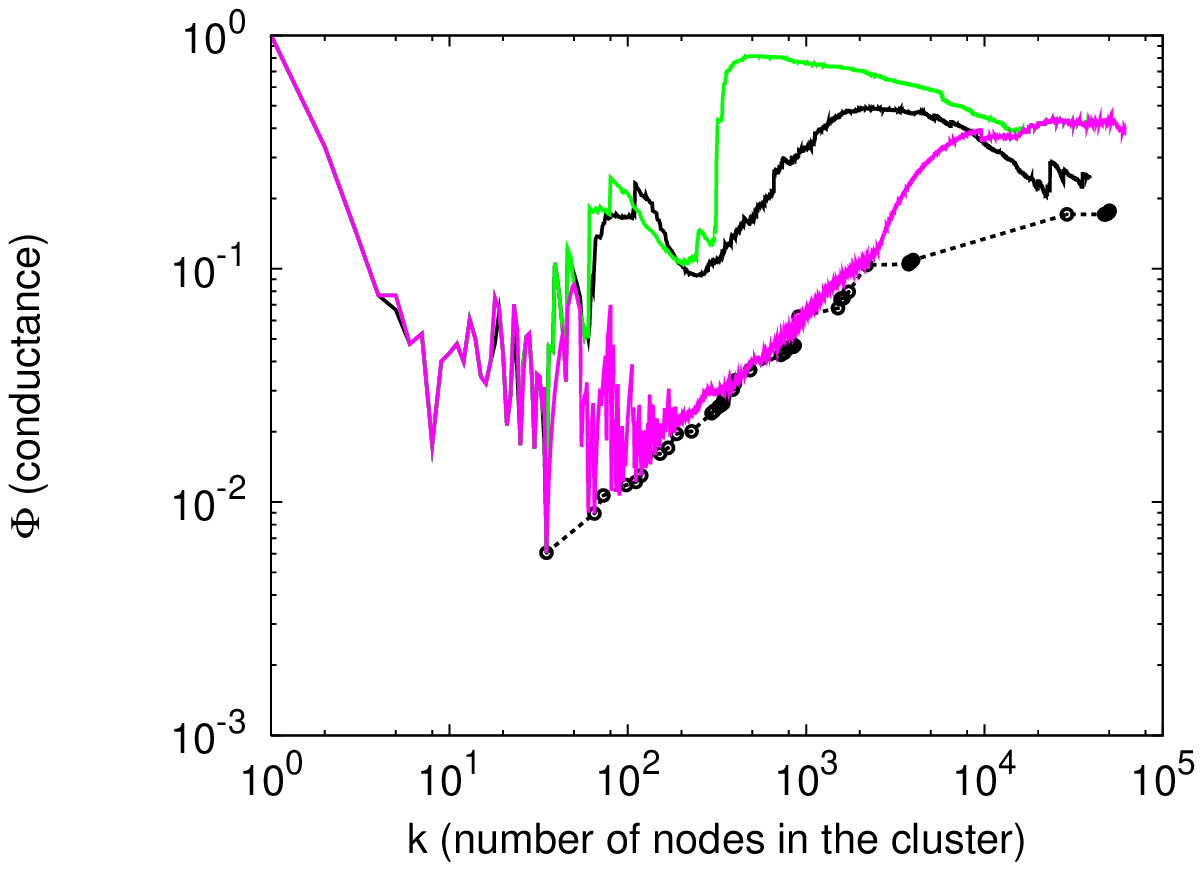}
    \includegraphics[width=0.32\linewidth]{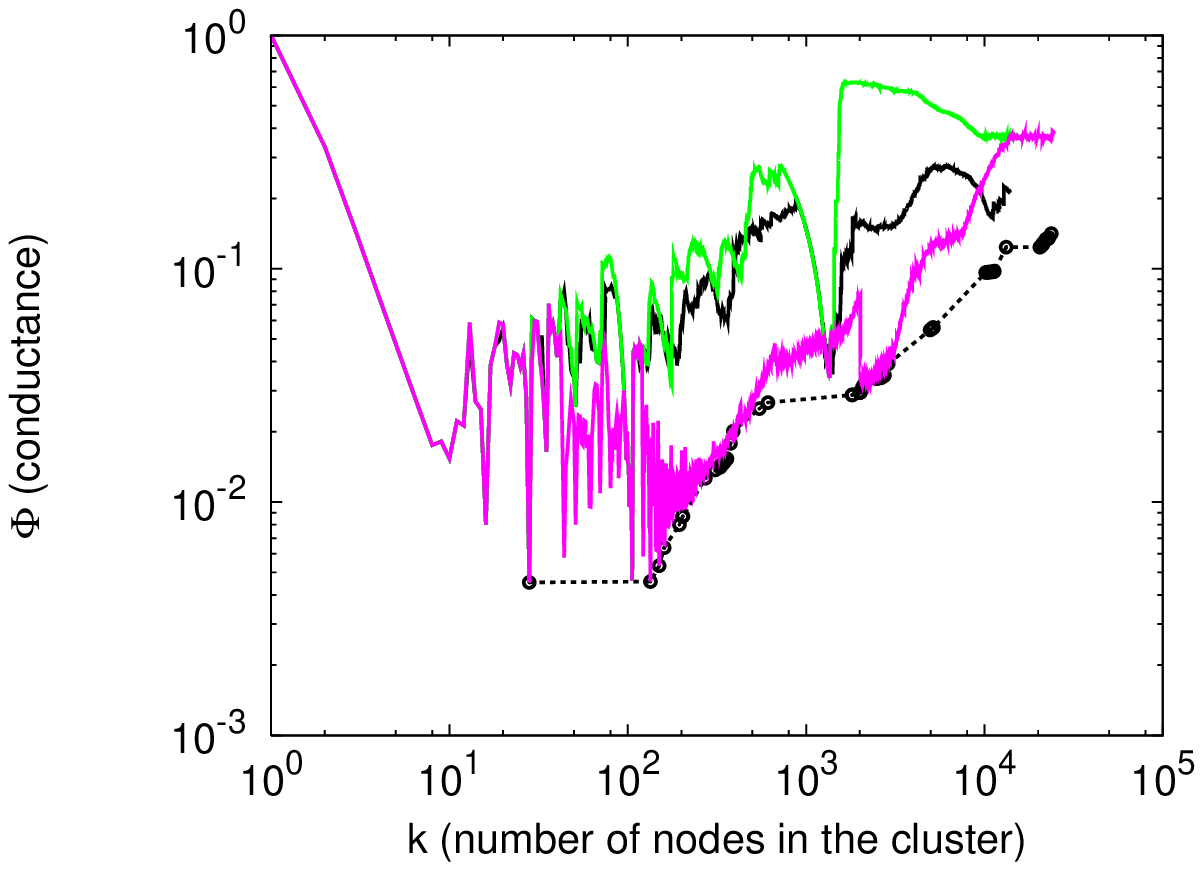}
    \includegraphics[width=0.32\linewidth]{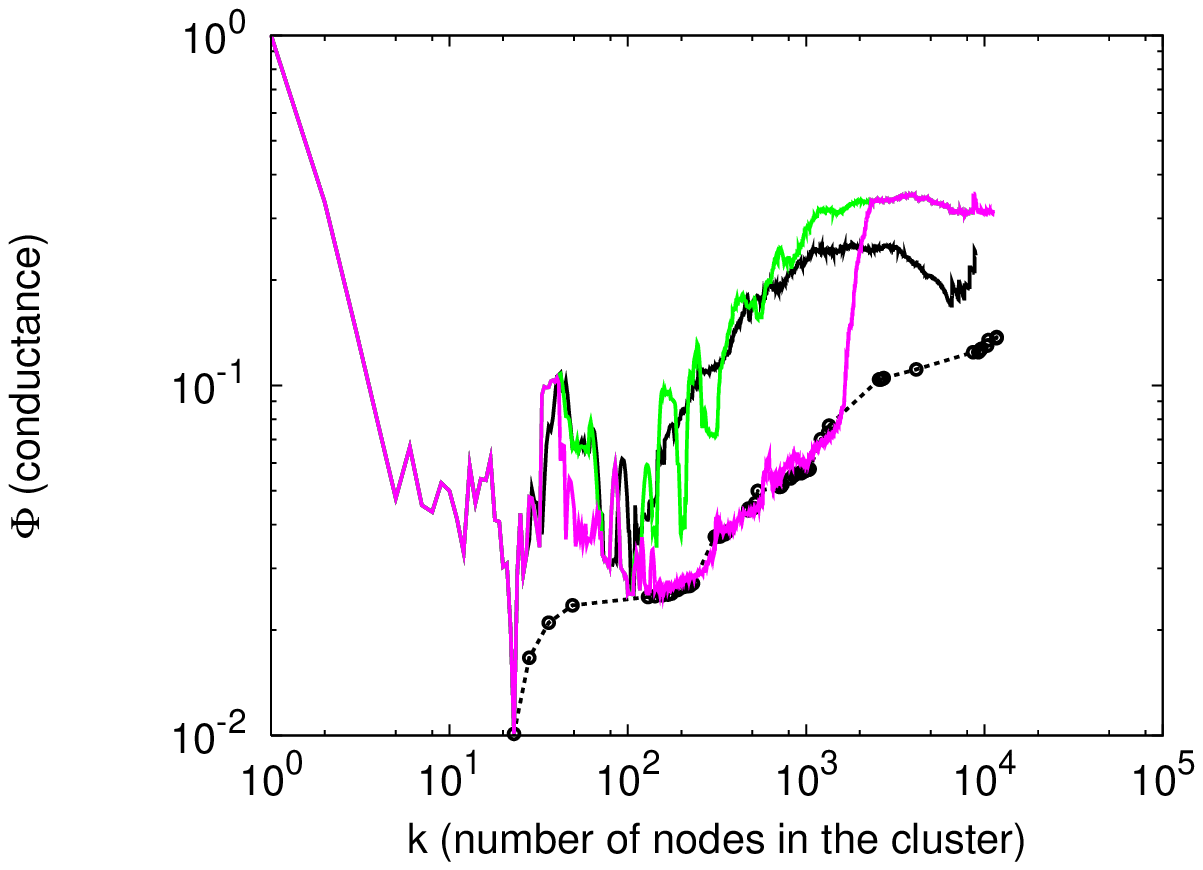}\\
    Leighton-Rao: connected clusters (green), disconnected clusters (magenta).\\
    \includegraphics[width=0.32\linewidth]{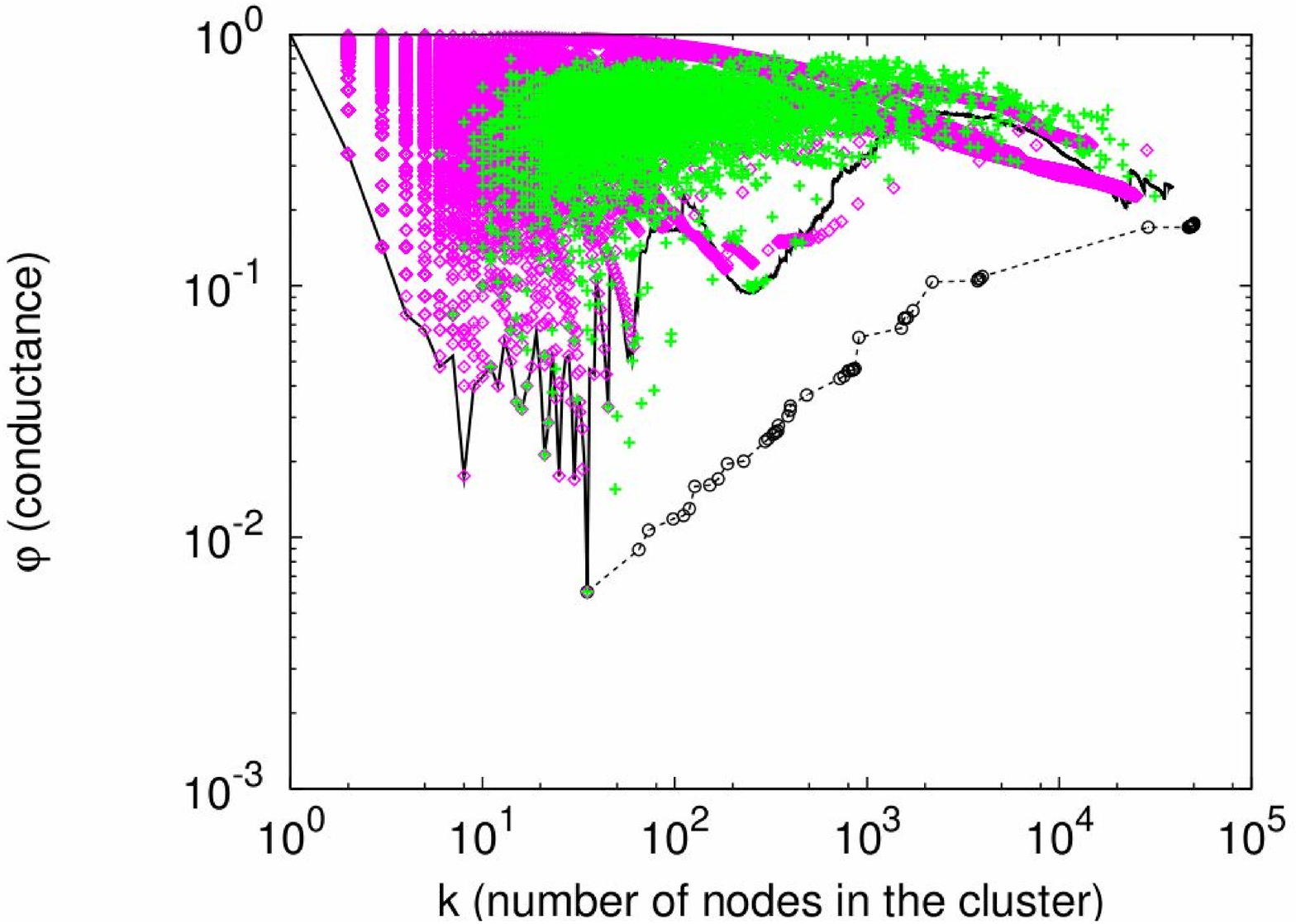}
    \includegraphics[width=0.32\linewidth]{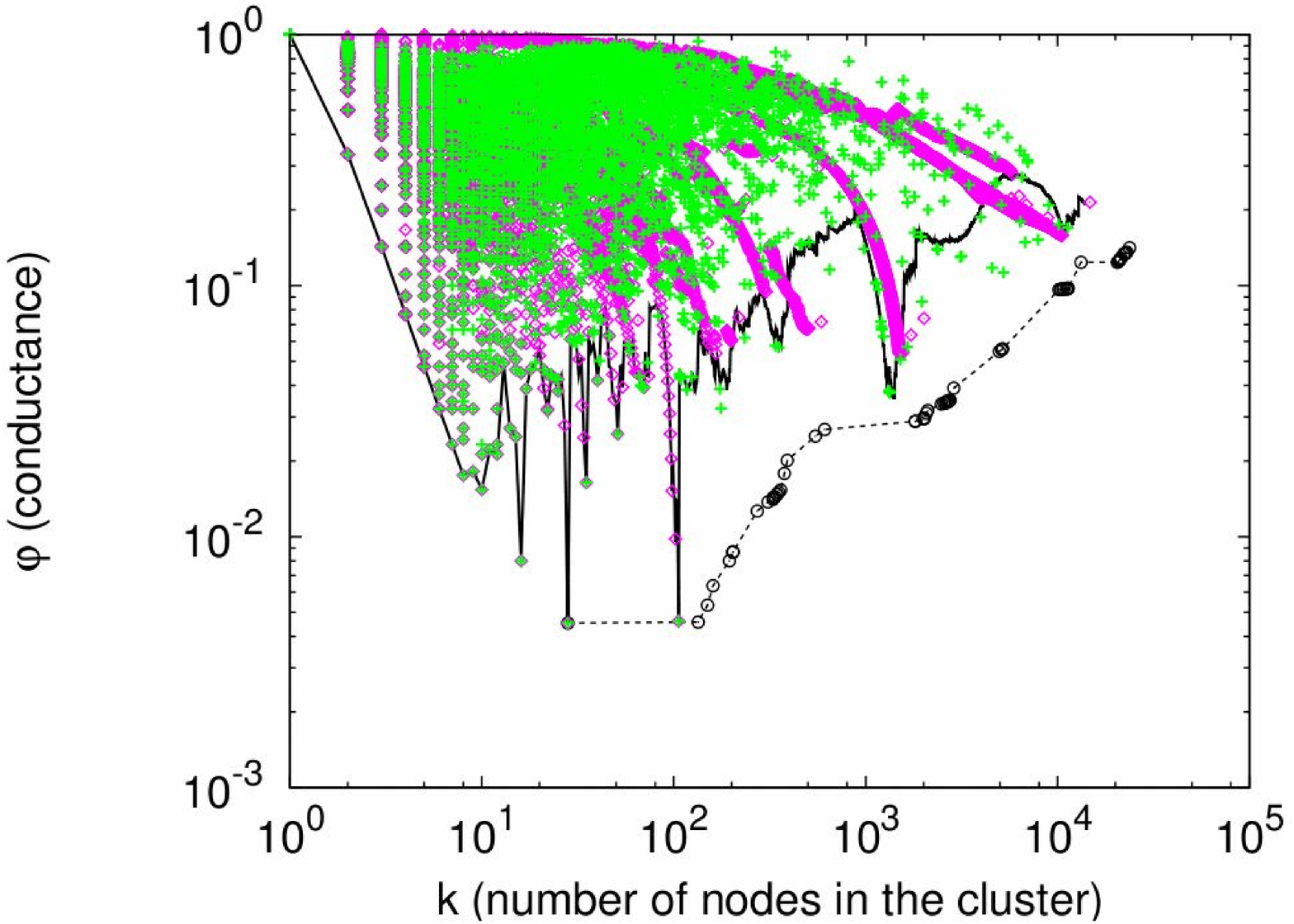}
    \includegraphics[width=0.32\linewidth]{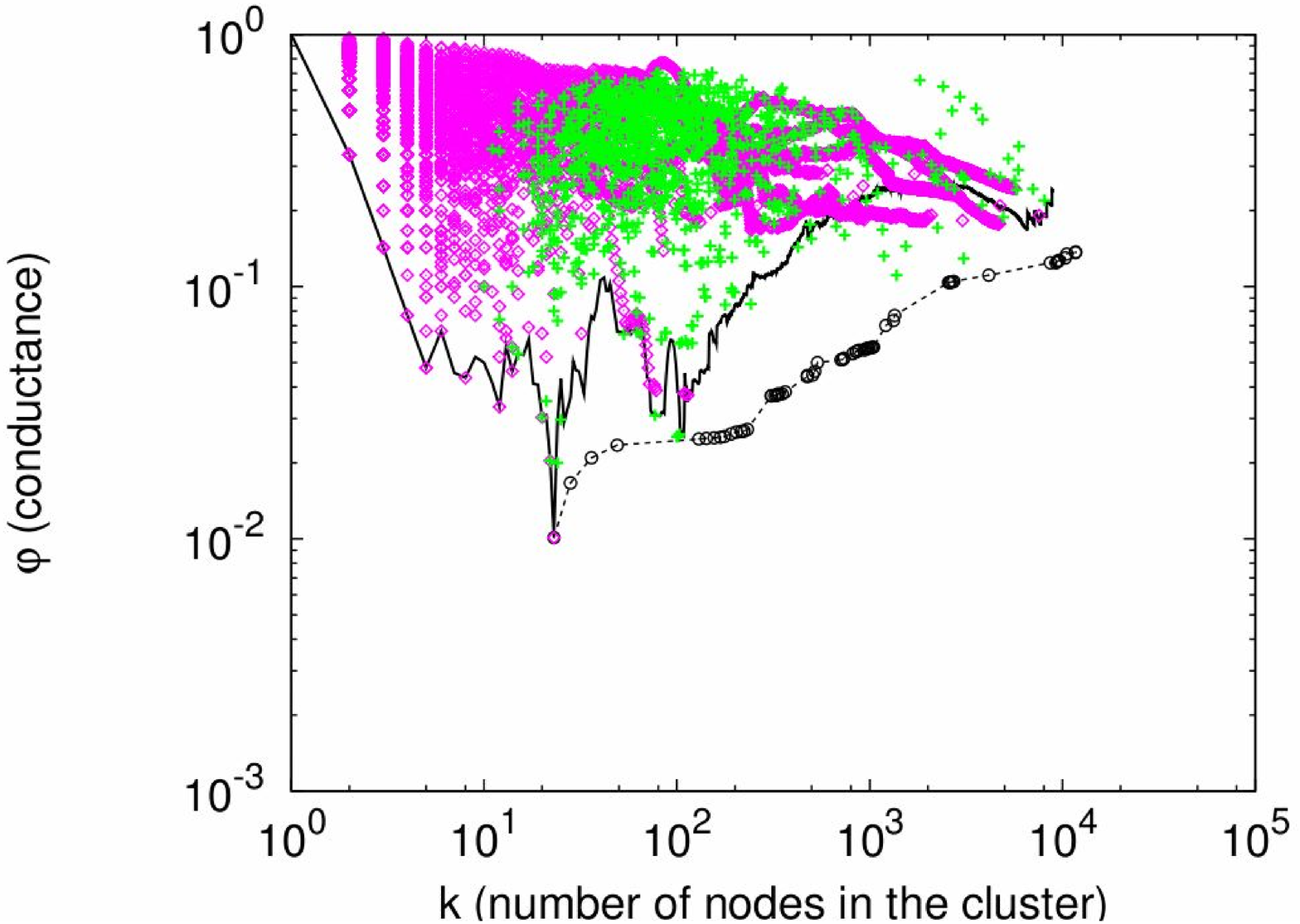}\\
    NCP plots obtained by Graclus and Newman's Dendrogram algorithm.\\
    \includegraphics[width=0.6\linewidth]{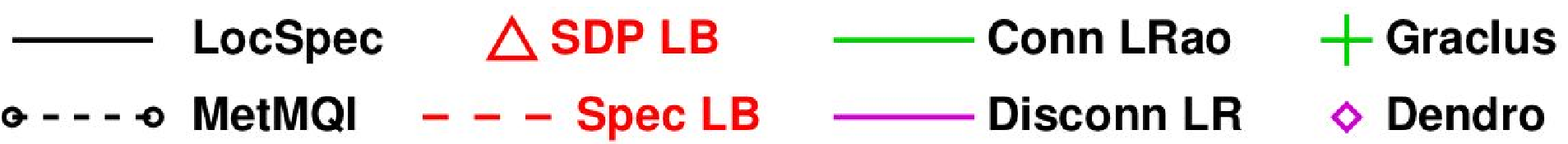}
\end{center}
\vspace{-6mm} \caption{ Comparison of various algorithms on \net{Epinions},
\net{Email-Enron}, and \net{CA-astro-ph}. Top row: NCP plots for connected
(green) and disconnected
(magenta) pieces from our implementation of the Leighton-Rao algorithm. %
Bottom row: conductance of cuts found by Graclus and by Newman's Dendrogram algorithm.
Notice the qualitative shape of the NCP plots remains practically unchanged
regardless of what particular community detection algorithm we use. Small
clusters tend to be connected, while for clusters larger than about 100 nodes
connected clusters have around 5 times worse conductance score than
disconnected clusters. } \vspace{-3mm} \label{other-algos-fig}
\end{figure*}

In top row of Figure~\ref{other-algos-fig}, we show Leighton-Rao curves
for three example graphs. Local Spectral and Metis+MQI curves are drawn in
black, while the Leighton-Rao curves for connected and possibly disconnected
sets are drawn in green and magenta respectively. 
For small to
medium scales, the Leighton-Rao curves for connected sets resemble the Local
Spectral curves, while the Leighton-Rao curves for possibly disconnected sets
resemble Metis+MQI curves. This further confirms the structure of
clusters produced by Local Spectral and Metis+MQI, as discussed in
Section~\ref{sec:pieces}.

At large scales, the Leighton-Rao curves shoot up and become much worse than
Local Spectral or Metis+MQI.  That Leighton-Rao has troubles finding
good big clusters is not surprising because expander graphs are known
to be the worst case input for the Leighton-Rao approximation guarantee.
Large real networks contain an expander-like core which is necessarily
encountered at large scales.
%%% COMBINE TWO PARS
We remark that Leighton-Rao does not work poorly at large scales on every
kind of graph. (In fact, for large low-dimensional mesh-like graphs,
Leighton-Rao is a very cheap and effective method for finding cuts at all
scales, while our local spectral method becomes impractically slow at medium
to large scales.) This means that based on the structure of the network and
sizes of clusters one is interested in different graph partitioning methods
should be used. While Leighton-Rao is an appropriate method for mesh-like graphs,
it has troubles in the intermingled expander-like core of large networks.

Finally, in addition to the above approximation algorithms-based methods for
finding low-conductance cuts, we also experimented with a number of more
heuristic approaches that tend to work well in practice. In particular, we
compare Graclus~\cite{dhillon07graclus} and Newman's modularity optimizing
program (we refer to it as Dendrogram)~\cite{newman02community}. Graclus
attempts to partition a graph into pieces bounded by
low-conductance cuts using a kernel $k$-means algorithm. We ran Graclus
repeatedly, asking for $2,3,\dots,i,i*\sqrt{2},...$ pieces. Then we
measured the size and conductance of all of the resulting pieces. Newman's
Dendrogram algorithm constructs a recursive partitioning of a graph (that is,
a dendrogram) from the bottom up by repeatedly deleting the surviving edge
with the highest betweenness centrality. A flat partitioning could then be
obtained by cutting at the level which gives the highest modularity score,
but instead of doing that, we measured the size of conductance of every piece
defined by a subtree in the dendrogram.

The bottom row of Figure~\ref{other-algos-fig} presents these results.  Again
our two standard curves are drawn in black. The lower-envelopes of the
Graclus or Dendrogram points are roughly similar to those produced by Local
Spectral, which means both methods tend to produce rather compact clusters at
all size scales. Generally, Graclus tends to produce a variety of clusters of
better conductance than Newman's algorithm. Moreover, notice that in case of
Epinions social network and the astrophysics coauthorship network Graclus
tends to prefer larger clusters than the Newman's algorithm. Also, Graclus seems to
find clusters of ten or more nodes, while Newmans's algorithm also extracts
very small pieces. In general, clusters produced by either Graclus or
Dendrogram are qualitatively similar to those produced by Local
Spectral. This means that even though Local Spectral is computationally
cheaper and easily scales to very large networks, the quality of identified
clusters 
is comparable to that returned by techniques such as Graclus
and Dendrogram that are significantly more expensive on large networks
such as those we considered.

\section{Comparison of objective functions}
\label{sec:scores}
In the previous sections, we used conductance since it corresponds most
closely to the intuition that a community is a set of nodes that is more and/or
better connected internally than externally. In this section, we look at
other objective functions that capture this intuition and/or are popular in
the community detection literature.

In general there are two criteria of interest when thinking about how good of
a cluster is a set of nodes. The first is the number of edges between the
members of the cluster, and the second is the number of edges between the members
of the cluster and the remainder of the network. We group objective
functions into two groups. The first group, that we refer to as Multi-criterion
scores, combines both criteria (number of edges inside and the number of edges
crossing) into a single objective function; while the second group of objective
functions employs only a single of the two criteria (\emph{e.g.}, volume of the
cluster or the number of edges cut).

\subsection{Multi-criterion scores}
\label{sxn:multi-crit-scores}

Let $G(V,E)$ be an undirected graph with $n=|V|$ nodes and $m=|E|$ edges. Let
$S$ be the set of nodes in the cluster, where $n_S$ is the number of nodes in
$S$, $n_S = |S|$; $m_S$ the number of edges in $S$, $m_S = |\{(u,v) : u \in
S, v\in S\}|$; and $c_S$, the number of edges on the boundary of $S$, $c_S =
|\{(u,v) : u \in S, v \not \in S\}|$; and $d(u)$ is the degree of node $u$.

We consider the following metrics $f(S)$ that capture the notion of a quality
of the cluster. Lower value of score $f(S)$ (when $|S|$ is kept constant)
signifies a more community-like set of nodes.
\begin{itemize}  \denselist
  \item {\bf Conductance:} $f(S)=\frac{c_S}{2 m_S + c_S}$ measures the
      fraction of total edge volume that points outside the
      cluster~\cite{ShiMalik00_NCut,kannan04_gbs}.
  \item {\bf Expansion:} $f(S)=\frac{c_S}{n_S}$ measures the number of
      edges per node that point outside the cluster~\cite{RCCLP04_PNAS}.
  \item {\bf Internal density:}  $f(S)=1 - \frac{m_S}{n_S (n_S-1)/2}$ is
      the internal edge density of the cluster $S$~\cite{RCCLP04_PNAS}.
  \item {\bf Cut Ratio:} $f(S)=\frac{c_S}{n_S (n - n_S)}$ is the fraction of
      all possible edges leaving the cluster~\cite{For09_TR}.
  \item {\bf Normalized Cut:} $f(S)=\frac{c_S}{2 m_S + c_S} +
      \frac{c_S}{2(m-m_S)+c_S}$~\cite{ShiMalik00_NCut}.
  \item {\bf Maximum-ODF (Out Degree Fraction):}\\$\max_{u \in S}
      \frac{|\{(u,v): v \not \in S\}|}{d(u)}$ is the maximum fraction of
      edges of a node pointing outside the
      cluster~\cite{flake00_efficient}.
  \item {\bf Average-ODF:} $f(S)=\frac{1}{n_S} \sum_{u \in S}
      \frac{|\{(u,v): v \not \in S\}|}{d(u)}$ is the average fraction of
      nodes' edges pointing outside the cluster~\cite{flake00_efficient}.
  \item {\bf Flake-ODF:} $f(S)=\frac{|\{u: u \in S, |\{(u,v): v \in S\}|
      < d(u)/2 \}|}{n_S}$ is the fraction of nodes in $S$ that have fewer edges
      pointing inside than to the outside of the
      cluster~\cite{flake00_efficient}.
\end{itemize}

We then generalize the NCP plot: for every cluster size $k$ we find a set of
nodes $S$ ($|S|=k$) that optimizes the chosen community score $f(S)$. We then
plot community score as a function of $k$. It is not clear how to design an
optimization procedure that would, given a cluster size $k$ and the community
score function $f(S)$, find the set $S$ that minimizes the function, \emph{i.e.}, is
the best community. Operationally, we perform the optimization the following
way: we use the Local Spectral method which starts from a seed node
and then explores the cluster structure around the seed node; running Local
Spectral from each node, we obtain a millions of sets of nodes of various
sizes, many of which are overlapping; and then for each such set of nodes, we
compute the community score $f(S)$ and find the best cluster of each size.

\begin{figure}[t!]
\begin{center}
\includegraphics[width=0.99\linewidth]{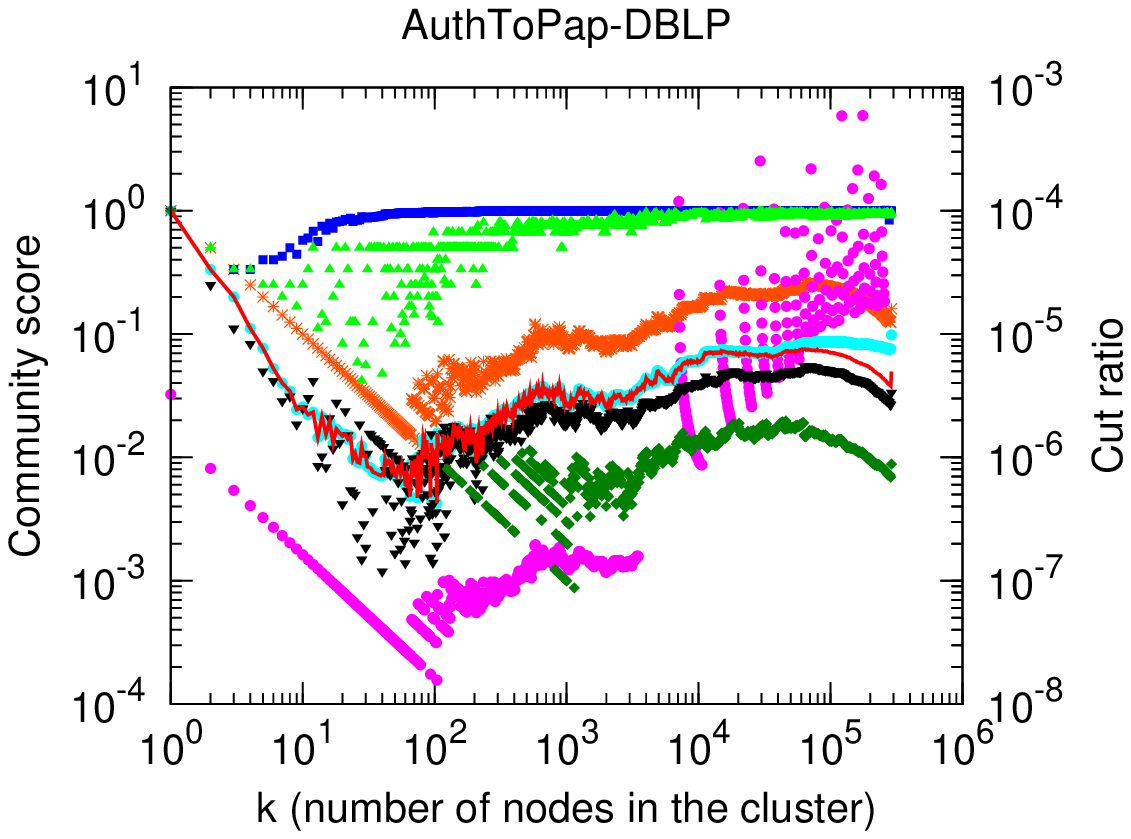} \\
\includegraphics[width=0.99\linewidth]{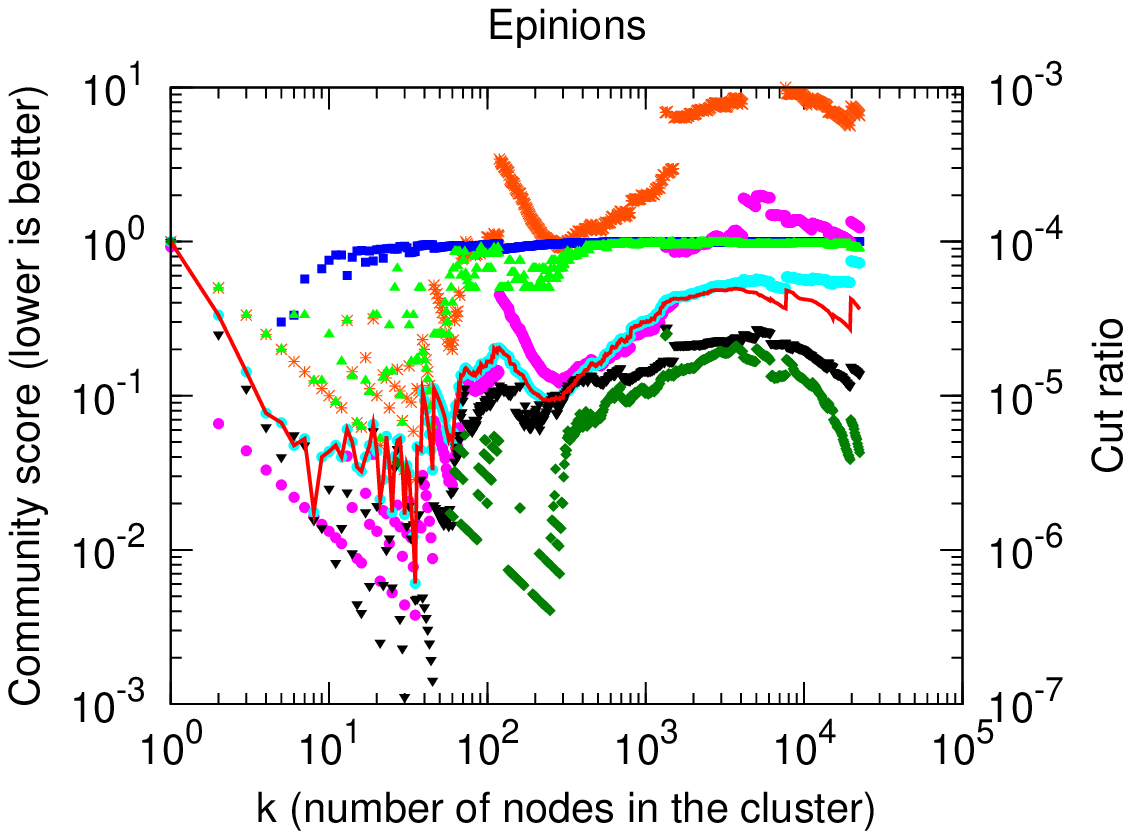} \\
\includegraphics[width=0.99\linewidth]{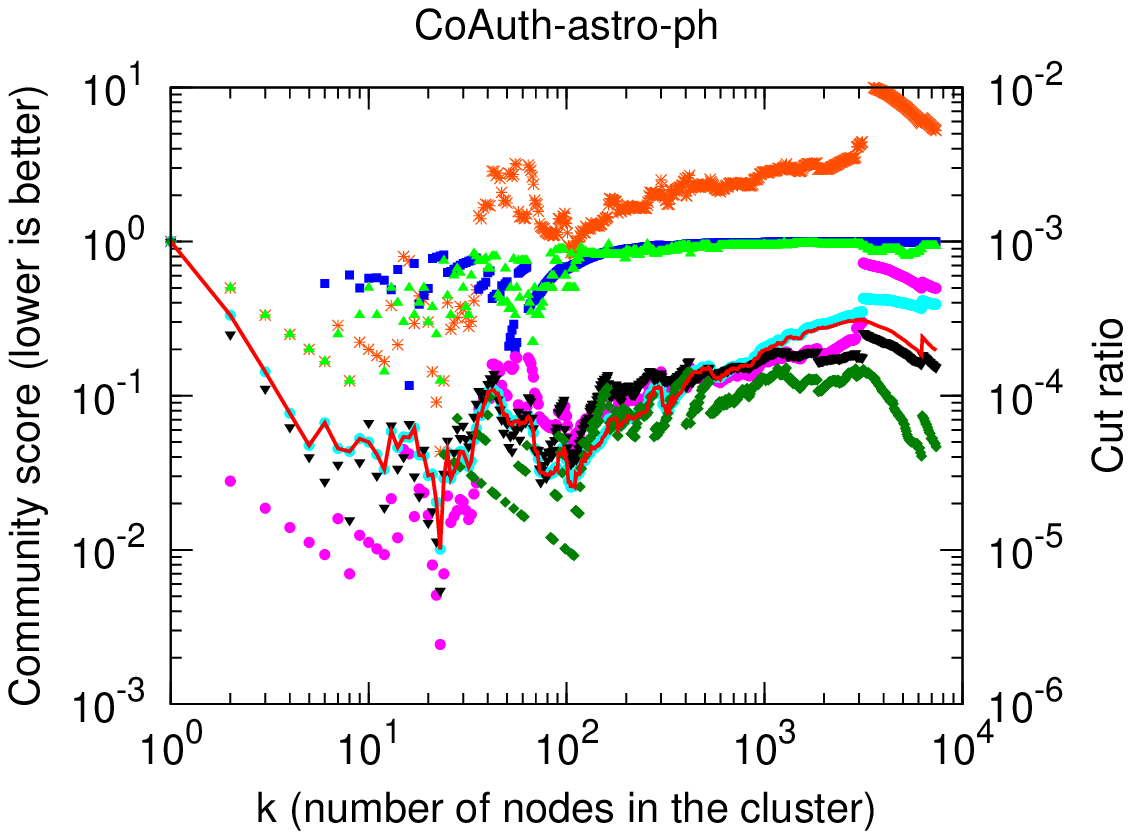}\\
\includegraphics[width=0.99\linewidth]{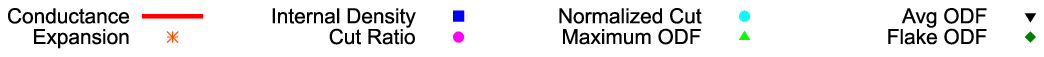}
\end{center}
\vspace{-5mm}
\caption{Various notions of community score as a function of
cluster size. All community scores (see main text for descriptions) have
qualitatively similar behaviors. They tend to decrease at first, which means
clusters get increasingly more community like as their size increases. After
that the score tends to degrade (it increases), which means clusters larger
than about 100 nodes get progressively less community like.} \vspace{-3mm}
\label{fig:cmtyMeasures1}
\end{figure}

Figure~\ref{fig:cmtyMeasures1} considers the above eight community scores.
Notice that even though scores span different ranges they all experience
qualitatively similar behavior, where clusters up to size ca. 100 have
progressively better scores, while the clusters above ca. 100 nodes become
less community-like as their size increases. This may seem surprising at
the first sight, but it should be somewhat expected, as all these objective
functions try to capture the same basic intuition---they reward sets of
nodes that have many edges internally and few pointing out of the clusters.

There are, however, subtle differences between various scores.  
For example, even though Flake-ODF follows same general trend as conductance, it
reaches the minimum about an order of magnitude later than conductance,
normalized cut, cut ratio score or the Average-ODF. On the other hand,
Maximum-ODF exhibits the opposite behavior as it clearly prefers smaller
clusters and is basically flat for clusters larger than about several hundred
nodes. This is interesting as this shows the following trend: if one scores
the community by the ``worst-case'' node using the Out Degree Fraction (\emph{i.e.},
Maximum-ODF) then only small clusters have no outliers and thus give good
scores. When one considers the average fraction of node's edges pointing
outside the cluster (Average-ODF) the objective function closely follows the
trend of conductance. On the other hand, if one considers the fraction of
nodes in the cluster with more of their edges pointing inside than outside
the cluster (Flake-ODF), then large clusters are preferred.

Next, focusing on the cut ratio score we notice that it is not very smooth, in the
sense that even for large clusters its values seem to fluctuate quite a lot.
This indicates that clusters of similar sizes can have very different
numbers of edges pointing to the rest of the network.  In terms of their
internal density, the variations are very small---the internal density
reaches the maximum for clusters of sizes around 10 nodes and then quickly
raises to 1, which means larger clusters get progressively sparser. For large
clusters this is not particularly surprising as the normalization factor
increases quadratically with the cluster size. This can be contrasted with
the Expansion score that measures the number of edges pointing outside the
cluster but normalizes by the number of nodes (not the number of all possible
edges).

These experiments suggest that Internal Density and Maximum-ODF
are not particularly good measures of community score and the cut ratio score
may not be preferred due to high variance. Flake-ODF seems to prefer larger
clusters, while conductance, expansion, normalized cut, and Average-ODF all
exhibit qualitatively similar behaviors and give best scores to similar
clusters.

In addition, we performed an experiment where we extracted clusters based on their
conductance score but then also computed the values of other community scores
(for these same clusters). This way we did not optimize each community score
separately, but rather we optimized conductance and then computed values of
other objective functions on these best-conductance pieces. The shape of the
plots remained basically unchanged, which suggests that same sets of nodes
achieve relatively similar scores regardless of which particular notion of
community score is used (conductance, expansion, normalized cut, or
Average-ODF). This shows that these four community scores are highly
correlated and in practice prefer practically the same clusters.

\subsection{Single criterion scores}

Next we also consider community scores that consider a single criteria. One
such example is Modularity~\cite{newman2006_ModularityPNAS}, which is one of
the most widely used methods to evaluate the quality of a division of a
network into modules or communities. For a given partition of a network into
clusters, modularity measures the number of within-community edges, relative
to a null model of a random graph with the same degree distribution.

Here we consider the following four notions of a quality of the community
that are based on using one or the other of the two criteria of the 
previous subsection:

\begin{itemize} \denselist
  \item {\bf Modularity:} $\frac{1}{4 m}(m_S - E(m_S))$, where $E(m_S)$
      is the expected number of edges between the nodes in set $S$ in a
      random graph with the same node degree sequence.
  \item {\bf Modularity ratio:} $\frac{m_S}{E(m_S)}$ is alternative
      definition of the modularity, where we take the ratio of the number
      of edges between the nodes of $S$ and the expected number of such
      edges under the null-model.
  \item {\bf Volume:} $\sum_{u \in S} d(u)$ is sum of degrees of nodes in
      $S$.
  \item {\bf Edges cut:} $c_S$ is number of edges needed to be removed to
      disconnect nodes in $S$ from the rest of the network.
\end{itemize}

\begin{figure*}[t!]
\begin{center}
\includegraphics[width=0.32\linewidth]{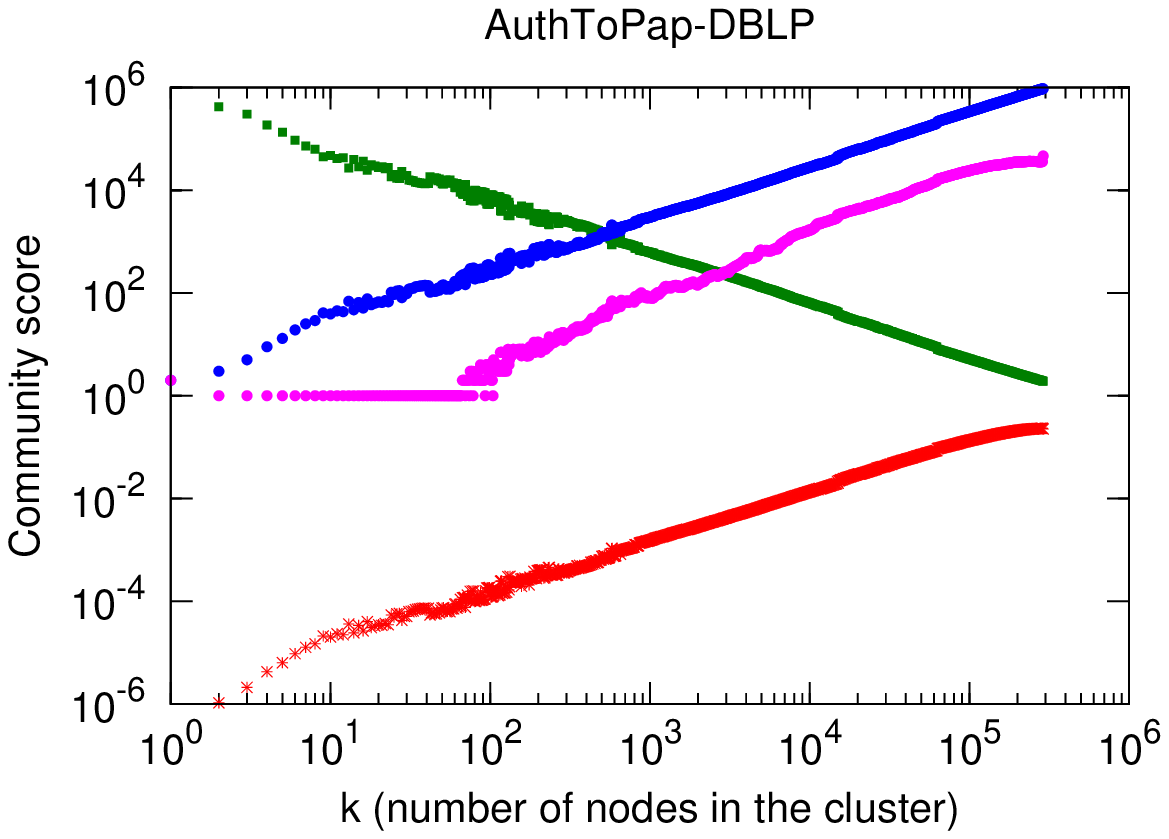}
\includegraphics[width=0.32\linewidth]{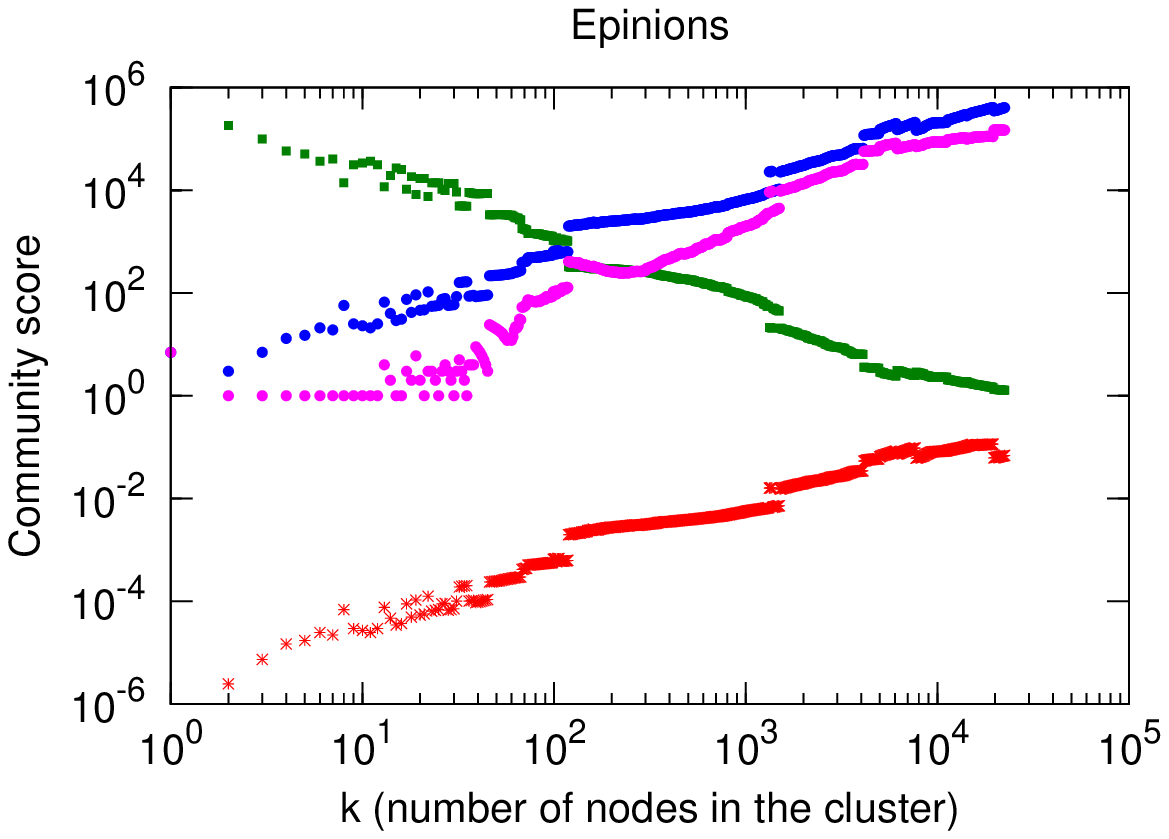}
\includegraphics[width=0.32\linewidth]{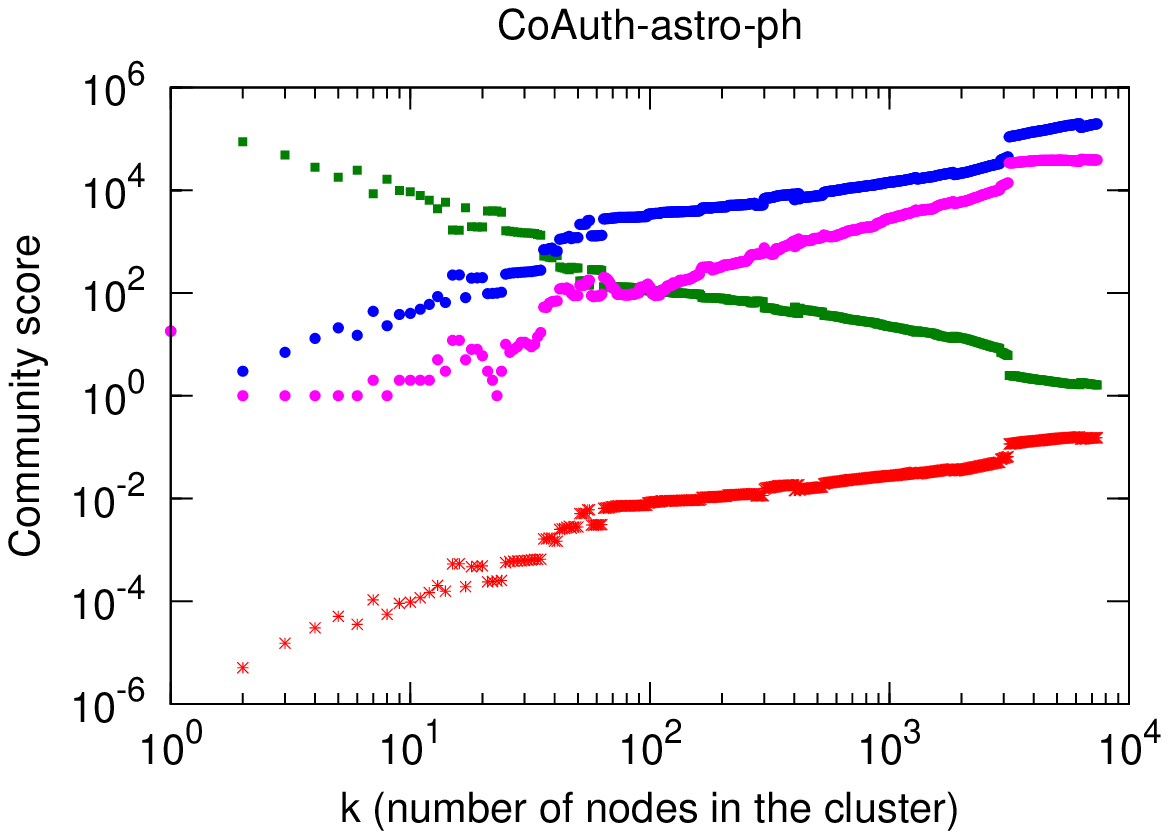}\\
\includegraphics[width=0.8\linewidth]{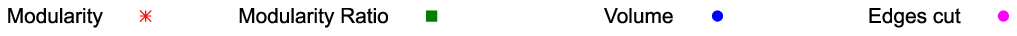}
\end{center}
\vspace{-5mm}
\caption{Four notions of community quality based on modularity.
The curves do not exhibit any particularly interesting non-monotonic trends.
Partitions of roughly half the network tend to have best modularity.
}
\vspace{-3mm}
\label{fig:cmtyMeasures2}
\end{figure*}

Figure~\ref{fig:cmtyMeasures2} shows the analog of the NCP plot where now
instead of conductance we use these four measures.
A general observation is that modularity
tends to increase roughly monotonically towards the bisection of the network.
This should not be surprising since modularity measures the ``volume'' of
communities, with (empirically, for large real-world networks) a small additive correction, and the volume clearly
increases with community size. On the other hand, the modularity ratio tends
to decrease towards the bisection of the network. This too should not be
surprising, since it involves dividing the volume by a relatively small
number.
%%%
Results in Figure~\ref{fig:cmtyMeasures2} demonstrate that, with respect to
the modularity, the ``best'' community in any of these networks has about half of
all nodes; while, with respect to the modularity ratio, the
``best'' community in any of these networks has two or three nodes. 

%%% In
%%% particular, 
Leaving aside debates about community-quality objective functions, note that,
whereas the conductance and related measures are \emph{discriminative},
in that they prefer different kinds of clusters, depending on the type of 
network being considered, modularity tends to follow
the same general pattern for all of these classes of networks.
That is, even aside from community-related interpretations,
conductance (as well as several of the other
bi-criterion objectives considered in Section~\ref{sxn:multi-crit-scores}) 
has qualitatively
different types of behaviors for very different types of graphs (\emph{e.g.},
low-dimensional graphs, expanders, large real-world social and
information networks), whereas modularity and other single-criterion
objectives behave in qualitatively similar ways for all these diverse classes
of graphs.

\section{Computing lower bounds}
\label{sec:bounds}
So far we have examined various heuristics and approximation algorithms for
community detection and graph partitioning. Common to these approaches is
that they all only approximately find good cuts, \emph{i.e.}, they only approximately
optimize the value of the objective function. Thus the clusters they identify
provide only an \emph{upper bound} on the true minimum best clusters. To get a better
idea of how good those upper bounds are, we compute theoretical \emph{lower bounds}.
Here we discuss the spectral lower bound~\cite{Chung:1997} on the conductance
of cuts of arbitrary balance, and a related SDP-based lower
bound~\cite{burer03lowrank} on the conductance of any cut that divides the
graph into two pieces of equal volume.

Lower bounds are usually not computed for practical
reasons, but instead are used to gain insights into partitioning
algorithms and properties of graphs where algorithms perform well or poorly. 
Also, note that the lower
bounds are ``loose,'' in the sense that they do not guarantee that
a cluster of a particular score exists; rather they are just saying that there
exists no cluster of better score.

First, we introduce the notation: $\vec{d}$ is a column vector of the graph's
node degrees; $D$ is a square matrix whose only nonzero entries are the
graph's node degrees on the diagonal; $A$ is the adjacency matrix of $G$;
$L=D-A$ is then the non-normalized Laplacian matrix of $G$; {\bf 1} is vector
of 1's; and $A \bullet B = trace (A^T B)$ is the matrix dot-product operator.
%%%
Now, consider the following optimization problem (which is well known to be
equivalent to an eigenproblem):
%%%
\[
\lambda_G =
  {\rm min}
  \left\{
    \frac{x^T L x}{x^T D x}: x \perp \vec{d}, x \neq 0
  \right\}   .
\]
%%%
Let $\hat{x}$ be a vector achieving the minimum value $\lambda_G$. Then
$\frac{\lambda_G}{2}$ is the spectral lower bound on the conductance of
any cut in the graph, regardless of balance, while $\hat{x}$ defines a
spectral embedding of the graph on a line, to which rounding algorithms
can be applied to obtain actual cuts that can serve as upper bounds at
various sizes.

Next, we discuss an SDP-based lower bound on cuts which partition the
graph into two sets of exactly equal volume. Consider:
%%%
\[
\mathcal{C}_G =
  {\rm min}
  \left\{
   \frac{1}{4}L\bullet Y: diag(Y) = {\bf 1}, Y\bullet(\vec{d}\,\vec{d}^{\,T}) = 0, Y \succeq 0
  \right\}   ,
\]
%%%
and let $\hat{Y}$ be a matrix achieving the minimum value $\mathcal{C}_G$.
Then $\mathcal{C}_G$ is a lower bound on the weight of any cut with
perfect volume balance, and $2 \mathcal{C}_G/{\rm Vol}(G)$ is a lower
bound on the conductance of any cut with perfect volume balance.  We
briefly mention that since $Y \succeq 0$, we can view $Y$ as a Gram matrix
that can be factored as $RR^T$.  Then the rows of $R$ are
the coordinates of an embedding of the graph on a hypersphere. Again,
rounding algorithms can be applied to the embedding to obtain actual cuts
that can serve as upper bounds.

\begin{figure*}[t!]
\begin{center}
\includegraphics[width=0.32\linewidth]{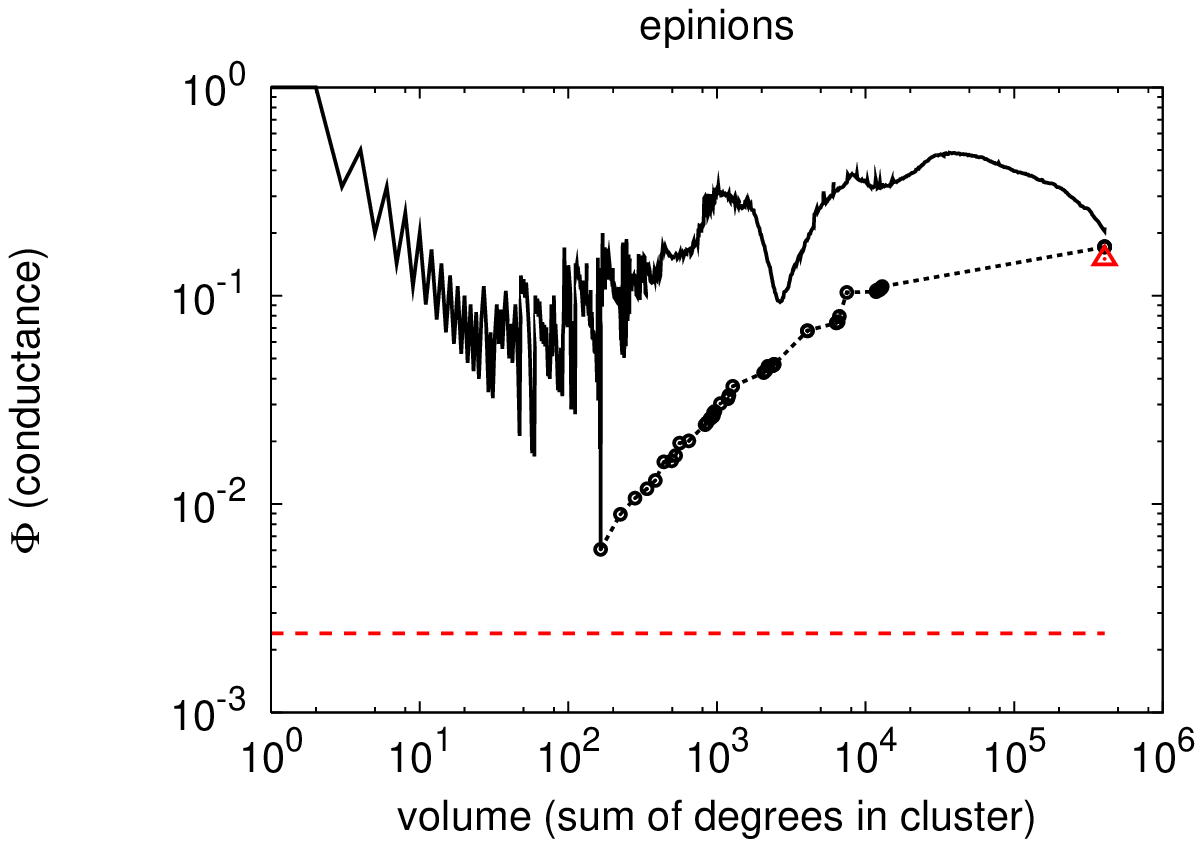}
\includegraphics[width=0.32\linewidth]{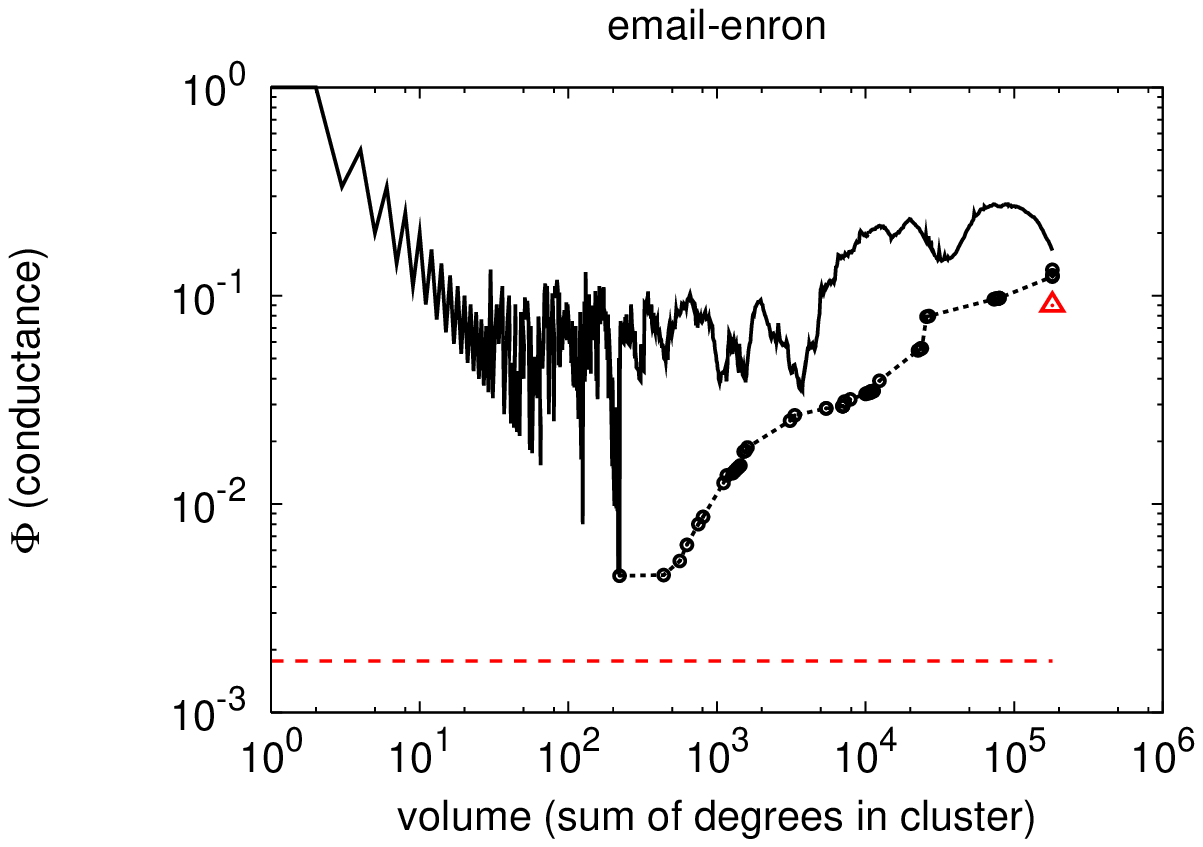}
\includegraphics[width=0.32\linewidth]{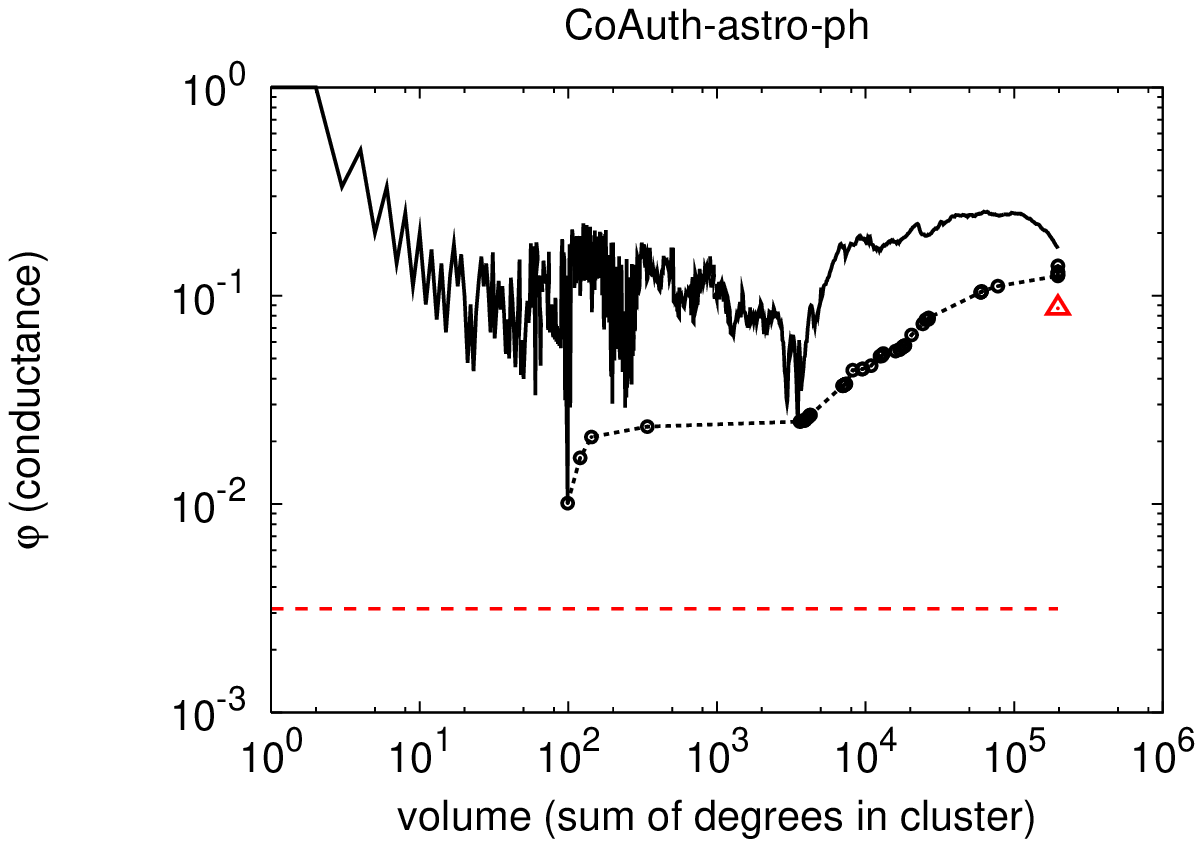}
\end{center}
\vspace{-6mm} \caption{Lower bounds on best conductance cut. In black we plot
conductance curves as obtained by Local Spectral and Metis+MQI. Lower bounds
on conductance of any cut (Spectral lower bound, dashed line) and the cut
separating the graph in half (SDP lower bound, red triangle) show that both
algorithms do a good job on finding best possible and best balanced cut of
the network. } \vspace{-2mm} \label{other-algos-fig-LBfig}
\end{figure*}

The spectral and SDP embeddings defined here were the basis for the extensive
experiments with global spectral partitioning methods that were alluded to in
Section~\ref{sec:algs}. In this section, it is the lower bounds that
concern us.
%%%
Figure~\ref{other-algos-fig-LBfig} shows the spectral and SDP lower bounds
for three example graphs. The spectral lower bound, which applies to cuts of
any balance, is drawn as a horizontal red line which appears near the bottom
of each plot. The SDP lower bound, which only applies to cuts separating a
specific volume, namely ${\rm Vol}(G)/2$, appears as an red triangle near the
right side of the each plot. (Note that plotting this point required us to
use volume rather than number of nodes for the x-axis of these plots.)

Clearly, for these graphs, the lower bound at ${\rm Vol}(G)/2$, is higher
than the spectral lower bound which applies at smaller scales. More
importantly, the lower bound at ${\rm Vol}(G)/2$, is higher than our {\em
upper} bounds at many smaller scales. This demonstrates two important points:
(1) It shows that best conductance clusters are orders of magnitude better
than best clusters consisting of half the edges; and (2) It demonstrates that
graph partitioning algorithms perform well at various size scales. For all
graph partitioning algorithms, the minimum of their NCP plot is close to the
spectral lower bound, and the clusters at half the volume are again close to
theoretically best possible clusters. This suggests that graph partitioning
algorithms we considered here do a good job both at finding best possible
clusters and at bisecting the network.

Take, for example, the first plot of Figure~\ref{other-algos-fig-LBfig}, where
in black we plot the conductance curves obtained by our (Local Spectral and
Metis+MQI) algorithms. With a red dashed line we plot the lower bound on the best
possible cut in the network, and with red triangle we plot the lower bound
for the cut that separates the graph in two equal volume parts. Thus, the
true conductance curve (which is intractable to compute) lies below black but
above red line and red triangle.
%%%
From practical perspective this demonstrates that the graph partitioning
algorithms (Local Spectral and Metis+MQI in particular) do a good job of
extracting clusters at all size scales. The lower bounds tell us that the
conductance curve which starts at upper left corner first has to go down and
reach the minimum close to the horizontal dashed line (Spectral lower bound)
and then sharply rise and ends up above the red triangle (SDP lower bound).
This verifies several things: 
(1) graph partitioning algorithms perform well
at all size scales, as the extracted clusters have scores close to the
theoretical optimum; 
(2) the qualitative shape of the NCP is not an artifact of graph partitioning
algorithms or particular objective functions, but rather it is an intrinsic property of these large networks;
and (3) the lower bounds at half the size of the graph indicate that
our inability to find large good-conductance communities is not a
failings of our algorithms.
Instead such large good-conductance ``communities'' 
simply do not exist in these networks.

Finally, in Table~\ref{lower-bound-table} we list for about 40 graphs the
spectral and SDP lower bounds on overall conductance and on volume-bisecting
conductance, and also the ratio between the two. It is interesting to see
that for these graphs this ratio of lower bounds does a fairly good job of
discriminating between declining-NCP-plot graphs, which have a small ratio,
and V-shape-NCP-plot graphs, which have a large ratio. Small networks (like
\net{CollegeFootball}, \net{ZacharyKarate} and \net{MonksNetwork}) have
downward NCP plot~\cite{LLDM08_communities_CONF,LLDM08_communities_TR} and a
small ratio of the SDP and Spectral lower bounds. On the other hand large
networks (\emph{e.g.}, \net{Epinions} or \net{Answers-3}) have downward and
then upward NCP plot (as in Figure~\ref{fig:intro}(right)) have large ratio
of the two lower bounds. This hints that in small networks ``large'' clusters
(i.e., clusters of around half the network) tend to have best conductances.
On the contrary, in large networks small clusters have good conductances,
while large clusters (of the half the network size) tend to have much worse
conductances, and thus high ratios of lower bounds as shown in the left table
of Table~\ref{lower-bound-table}\footnote{See~\cite{LLDM08_communities_TR}
for the NCPs of the networks listed in Table 1.}.

\begin{table*}[t!]
\begin{center}
{ \footnotesize
\subtable{
\begin{tabular}{|l|r|r|r|}
\hline
%%%Network & Spectral LB & SDP LB & Ratio \\
                    & Spectral   & SDP              &  ratio   \\
                    & lowerbnd   & lowerbnd         &  of      \\
                    & on $\phi$, & on $\phi$, at    &  lower   \\
            Network & any size.  & ${\rm Vol}(G)/2$ &  bnds    \\
\hline \hline
\net{CollegeFootball}~\cite{newman_netdata}
                    & 0.068402  &  0.091017  &  1.330624  \\
\net{MonksNetwork}~\cite{newman_netdata}
                    & 0.069660  &  0.117117  &  1.681269  \\
\net{ZacharyKarate}~\cite{newman_netdata}
                    & 0.066136  &  0.127625  &  1.929736  \\
\net{PowerGrid}     & 0.000136  &  0.000268  &  1.978484  \\
\net{PoliticalBooks}~\cite{newman_netdata}
                    & 0.018902  &  0.038031  &  2.011991  \\
\net{PoliticalBlogs}~\cite{newman_netdata}
                    & 0.040720  &  0.084052  &  2.064157  \\
%%%LesMiserables~\cite{newman_netdata} & 0.044067  &  0.099292  &  2.253198  \\
\net{RB-Hierarchical}~\cite{ravasz03_hierarchical}
                    & 0.011930  &  0.030335  &  2.542792  \\
\net{Email-InOut}   & 0.038669  &  0.113367  &  2.931752  \\
\net{NetworkScience}~\cite{newman_netdata}
                    & 0.001513  &  0.004502  &  2.974695  \\
\net{As-Oregon}     & 0.012543  &  0.042976  &  3.426417  \\
\net{Blog-nat05-6m} & 0.031604  &  0.108979  &  3.448250  \\
\net{Imdb-India}    & 0.009104  &  0.033318  &  3.659573  \\
\net{Cit-hep-ph}    & 0.007858  &  0.029243  &  3.721553  \\
\net{Bio-Proteins}  & 0.033714  &  0.126137  &  3.741358  \\
\net{As-RouteViews} & 0.018681  &  0.070462  &  3.771821  \\
\net{Gnutella-31}   & 0.029946  &  0.118711  &  3.964127  \\
\net{Imdb-Japan}    & 0.003327  &  0.013396  &  4.026721  \\
\net{Gnutella-30}   & 0.030621  &  0.124929  &  4.079853  \\
\net{DolphinsNetwork}~\cite{newman_netdata}
                    & 0.019762  &  0.103676  &  5.246171  \\
\net{As-Newman}     & 0.009681  &  0.058952  &  6.089191  \\
\net{AtP-gr-qc}     & 0.000846  &  0.006040  &  7.141270  \\
\net{Cit-hep-th}    & 0.009193  &  0.068880  &  7.492522  \\
\net{AtP-cond-mat}  & 0.001703  &  0.013452  &  7.897650  \\
\hline
\end{tabular}
} % end subtable
\subtable{
\begin{tabular}{|l|r|r|r|}
\hline
%%%Network & Spectral LB & SDP LB & Ratio \\
                    & Spectral   & SDP              &  ratio   \\
                    & lowerbnd   & lowerbnd         &  of      \\
                    & on $\phi$, & on $\phi$, at    &  lower   \\
            Network & any size.  & ${\rm Vol}(G)/2$ &  bnds    \\
\hline \hline
\net{Gnutella-25}   & 0.014185  &  0.131032  &  9.237332  \\
\net{Answers-2}     & 0.009660  &  0.107422  &  11.120081  \\
\net{CA-cond-mat}   & 0.003593  &  0.047064  &  13.098027  \\
\net{Answers-1}     & 0.011896  &  0.159251  &  13.386528  \\
\net{Imdb-France}   & 0.003462  &  0.048010  &  13.867591  \\
\net{Answers-5}     & 0.008714  &  0.124703  &  14.311255  \\
%%%airports            & 0.013701  &  0.211092  &  15.406824  \\
\net{Imdb-Mexico}   & 0.003893  &  0.070345  &  18.067513  \\
\net{CA-gr-qc}      & 0.000934  &  0.017421  &  18.659710  \\
\net{AtP-hep-th}    & 0.000514  &  0.009714  &  18.899660  \\
\net{AtP-hep-ph}    & 0.000723  &  0.013770  &  19.040287  \\
\net{Imdb-WGermany} & 0.003025  &  0.065158  &  21.538867  \\
\net{AtP-astro-ph}  & 0.001183  &  0.027256  &  23.036835  \\
\net{CA-hep-th}     & 0.001561  &  0.041125  &  26.350412  \\
\net{CA-astro-ph}   & 0.003143  &  0.086890  &  27.648094  \\
\net{Imdb-UK}       & 0.001283  &  0.036572  &  28.514376  \\
\net{Imdb-Germany}  & 0.000661  &  0.021017  &  31.810460  \\
\net{Blog-nat06all} & 0.002361  &  0.092908  &  39.350874  \\
\net{Imdb-Italy}    & 0.000679  &  0.031954  &  47.077242  \\
\net{Email-Enron}   & 0.001763  &  0.089876  &  50.965424  \\
\net{CA-hep-ph}     & 0.000889  &  0.052249  &  58.755927  \\
\net{Epinions}      & 0.002395  &  0.150242  &  62.739252  \\
\net{Answers-3}     & 0.002636  &  0.185340  &  70.306807  \\
\net{Imdb-Spain}    & 0.000562  &  0.046327  &  82.397702  \\
\hline
\end{tabular}
} % end subtable
} % end footnotesize
\end{center}
\vspace{-7mm} \caption{ Lower bounds on the conductance for our network
datasets. Spectral lower bound applies to any cut, while the SDP lower bound
applies to cuts at a specified volume fraction, taken here to be half. As
small networks have low ratio of lower bounds, large networks have much
higher ratios which demonstrates that in large networks good clusters tend to
be relatively small. Description and the statistics of the networks are
available at supporting website: {\tt http://snap.stanford.edu/ncp}. }
\vspace{-3mm} \label{lower-bound-table}
\end{table*}

\section{Conclusion}
\label{sec:conclusion}
In this paper we examined in a systematic way a wide range of network
community detection methods originating from theoretical computer science,
scientific computing, and statistical physics. Our empirical results 
demonstrate that determining the clustering structure of large networks is
surprisingly intricate. In general, algorithms nicely optimize the community
score function over a range of size scales, and the scores of obtained
clusters are relatively close to theoretical lower bounds. However, there are
classes of networks where certain algorithms perform sub-optimally. In
addition, although many common community quality objectives tend to exhibit
similar qualitative behavior, with very small clusters achieving the best
scores, several community quality metrics such as the commonly-used
modularity behave in qualitatively different ways.

Interestingly, intuitive notions of cluster quality tend to fail as one
aggressively optimizes the community score. For instance, by aggressively
optimizing conductance, one obtains disconnected or barely-connected clusters
that do not correspond to intuitive communities. 
This suggests the rather interesting point (that we described in 
Section~\ref{sec:pieces}) that \emph{approximate} optimization of the 
community score introduces a systematic bias into the extracted clusters, 
relative to the combinatorial optimum. Many times, as in case
of Local Spectral, such bias is in fact preferred since the resulting
clusters are more compact and thus correspond to more intuitive communities.
This connects very nicely to regularization concepts in machine learning and
data analysis, where separate penalty terms are introduced in order to
trade-off the fit of the function to the data and its smoothness. In our case
here, one is trading off the conductance of the bounding cut of the cluster
and the internal cluster compactness. Effects of regularization by
approximate computation are pronounced due to the extreme sparsity of real
networks. 
How to formalize a notion of regularization by approximate computation more 
generally is an intriguing question raised by our findings.

%\section*{Acknowledgments}
%MWM was supported in part by a grant from the AFOSR.
%JL was supported in part by a gift from Microsoft.

\bibliographystyle{abbrv}
%%%\bibliography{empirical}
%\bibliography{communities2}

\end{document}